\documentclass[sigconf,authorversion,nonacm]{acmart}
\pdfoutput=1
\usepackage{cite}
\usepackage{amsmath,amsfonts}
\usepackage{algorithmic}
\usepackage{graphicx}
\usepackage{textcomp}
\usepackage{xcolor}
\usepackage{fancyhdr}
\usepackage{multirow}
\usepackage{enumitem}
\usepackage{nccmath}
\usepackage{tabularx}
\usepackage{adjustbox}

\def\BibTeX{{\rm B\kern-.05em{\sc i\kern-.025em b}\kern-.08em
    T\kern-.1667em\lower.7ex\hbox{E}\kern-.125emX}}
\usepackage{pifont}
\newcommand\tab[1][1cm]{\hspace*{#1}}

\newcommand{\fixme}[1]{{\textcolor{black}{#1}}}
\newcommand{\wip}[1]{{\textcolor{magenta}{[WIP].}}} 
\newcommand{\co}{$\textrm{CO}_{2e}$}
\newcommand{\cdp}{\textcolor{black}{tCDP}}

\usepackage{soul}

\begin{document}

\title{Design Space Exploration and Optimization for Carbon-Efficient Extended Reality Systems}

\author{Mariam Elgamal}
\email{mariamelgamal@g.harvard.edu}
\affiliation{
  \institution{Harvard University/Meta}
   \country{USA}
}

\author{Doug Carmean}
\email{carmean@meta.com}
\affiliation{
  \institution{Meta}
   \country{USA}
}

\author{Elnaz Ansari}
\email{elnazans@meta.com}
\affiliation{
  \institution{Meta}
   \country{USA}
}

\author{Okay Zed}
\email{okay@meta.com}
\affiliation{
  \institution{Meta}
   \country{USA}
}

\author{Ramesh Peri}
\email{rvperi@meta.com}
\affiliation{
  \institution{Meta}
   \country{USA}
}

\author{Srilatha Manne}
\email{bmanne@meta.com}
\affiliation{
  \institution{Meta}
   \country{USA}
}

\author{Udit Gupta}
\email{uditg@meta.com}
\affiliation{
  \institution{Meta}
   \country{USA}
}

\author{Gu-Yeon Wei}
\email{guyeon@seas.harvard.edu}
\affiliation{
  \institution{Harvard University}
   \country{USA}
}

\author{David Brooks}
\email{dbrooks@eecs.harvard.edu}
\affiliation{
  \institution{Harvard University/Meta}
     \country{USA}
}

\author{Gage Hills}
\email{ghills@seas.harvard.edu}
\affiliation{
  \institution{Harvard University}
   \country{USA}
}

\author{Carole-Jean Wu}
\email{carolejeanwu@meta.com}
\affiliation{
  \institution{Meta}
   \country{USA}
}


\thispagestyle{firstpage}
\pagestyle{plain}

\begin{abstract}
As computing hardware becomes more specialized, designing environmentally sustainable computing systems requires accounting for both hardware and software parameters. Our goal is to design low carbon computing systems while maintaining a competitive level of performance and operational efficiency. Despite previous carbon modeling efforts for computing systems, there is a distinct lack of holistic design strategies to simultaneously optimize for carbon, performance, power and energy. In this work, we take a data-driven approach to characterize the carbon impact (quantified in units of \co) of various artificial intelligence (AI) and extended reality (XR) production-level hardware and application use-cases. We propose a holistic design exploration framework to optimize and design for carbon-efficient computing systems and hardware. Our frameworks identifies significant opportunities for carbon efficiency improvements in application-specific and general purpose hardware design and optimization. Using our framework, we demonstrate 10$\times$ carbon efficiency improvement for specialized AI and XR accelerators (quantified by a key metric, \cdp: the product of total \co~and total application execution time), up to 21\% total life cycle carbon savings for existing general-purpose hardware and applications due to hardware over-provisioning, and up to 7.86$\times$ carbon efficiency improvement using advanced 3D integration techniques for resource-constrained XR systems.
\end{abstract}

\maketitle

\section{Introduction}

The world's push toward an environmentally sustainable society is highly dependent on the semiconductor industry, due to global-scale carbon footprints of sources such as electric vehicles to augmented and virtual reality (XR)\footnote{Augmented reality (AR), virtual reality (VR) and extended reality (XR) for the combination of both AR and VR.} wearable devices. As computing technology and digital accessibility proliferate, the carbon footprint (quantified in units of \co) of the Information and Communication Technology (ICT) sector is expected to exceed its current $1.2-2.4\%$ of global emissions~\citep{Andrae2015, Freitag2021}. In 2022 alone, digital adoption increased with 192 million more users connected to the Internet, 85 million more unique mobile phone users, and 300 additional, newly-announced hyperscale datacenters expected to be in operation by 2024~\citep{projection_users, projection_datacenter}. For the ICT sector to achieve an environmentally sustainable carbon footprint, computing system developers must optimize and  low-carbon systems for key application use cases, such as AI and XR, across different layers of the computing stack. 

Optimizing hardware for carbon efficiency leads to trade-offs from performance and energy optimizations. For many of today's computing systems, the dominating source of carbon footprint has shifted from \textit{operational carbon}: \co~due to energy consumption during technology or application use, to \textit{embodied carbon}: \co~from hardware manufacturing and production~\citep{chasingGupta2021}. The total life cycle carbon emissions of a computing system is the total of the system's embodied and operational carbon. The growing shift from operational to embodied carbon is the result of decades of operational energy efficiency optimization from integrated circuits. In some cases, embodied carbon accounts for 50\% of cloud computing's carbon footprint, and over 70\% of consumer electronics carbon footprint~\citep{chasingGupta2021}.

\begin{figure}[t]
\centering
\includegraphics[width=0.9\columnwidth]{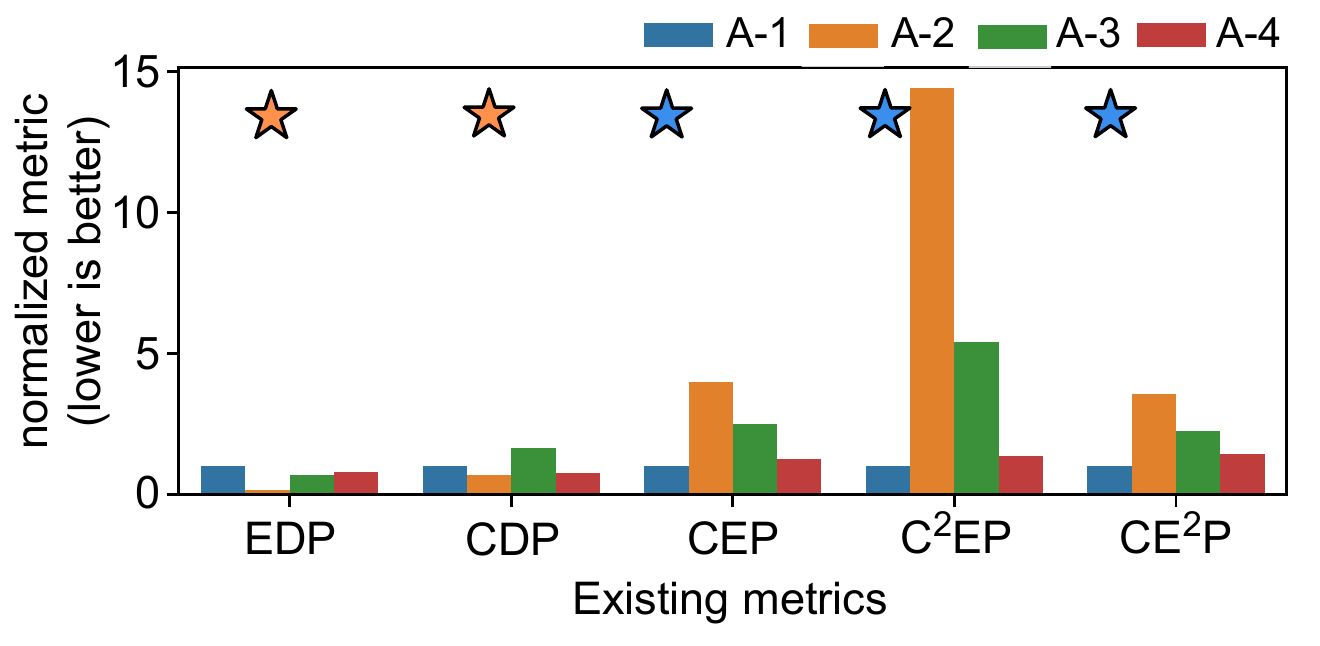}
\vspace{-0.45cm}
\caption{Accelerator A-2 is EDP and CDP optimal; A-1 is CEP, CE$^2$P, and C$^2$EP optimal. Existing state-of-the-art metrics do not capture the interplay between operational \textit{and} embodied carbon across the system's operational lifetime.
}
\label{fig:existing-metrics-accelerator} 
\vspace{-1em}
\end{figure}

Prior carbon accounting efforts quantify carbon emissions of computing systems from  semiconductor manufacturing and hardware production to operational use and end-of-life processing~\citep{TeehanLCA, boyd2011thesis, AJones_consideringfabrication, AJones_greencomputinglca}. Carbon modeling tools, such as GreenChip~\citep{Greenchip2019} and the architectural carbon modeling tool ACT~\citep{ACTGupta2022}, enable carbon-aware design exploration. ACT proposes carbon metrics such as embodied carbon-delay product (CDP) to optimize for embodied carbon and performance, and embodied carbon-energy product (CEP) to optimize for embodied carbon and energy. Additionally, recent work uses proxies to provide intuitions into systems' carbon footprint in the absence of industry carbon databases at design time~\citep{eeckhout2022}. These efforts are a first step to model system hardware's carbon footprint and explore carbon-aware systems.

Figure~\ref{fig:existing-metrics-accelerator} presents the results of EDP, CDP, CEP, C$^2$EP, CE$^2$P for four real-production hardware accelerators, A-1 to A-4. The optimal accelerator for the applications under study based on EDP and the state-of the-art sustainability metrics---\textit{CDP, CEP, CE$^2$P, C$^2$EP}---differs. Accelerator A-2 is EDP and CDP optimal due to its higher compute and SRAM capacities, leading to higher performance and energy efficiency. Accelerator A-1 is CEP, CE$^2$P, and C$^2$EP optimal due to its low embodied carbon---approximately 4 times lower than A-2's and 3 times lower than A-3's embodied carbon. 

Here Figure 1 highlights the importance of choosing the right metric during system design. Even in this simple example, choosing to optimize for CDP versus CEP results in an entirely different design choice (i.e., A-2 vs. A-1); in Section 3, we justify the effectiveness of a specific metric --- \cdp~(the product of total \co~and total application execution time) --- to guide designers toward making carbon-efficient design decisions.

In this paper, we aim to address three outstanding deficiencies in today's carbon-aware system design: 
\begin{itemize}
    \item Carbon modeling tools must consider the changing relative ratio between embodied \textit{and} operational carbon based on application requirements, renewable energy availability, operational use and lifetime, and hardware design knobs.
    \item Hardware design tools should enable computer architects and designers to assess the impact of varying hardware design on performance, power, energy, and total life cycle emissions, i.e. operational \textit{and} embodied carbon.
    \item Hardware design tools should consider the role of workloads and applications, leading to varying hardware utilization (such as thread-level parallelism). This is to provision hardware efficiently at the design phase and minimize unused hardware real estate incurring higher total life cycle carbon.
\end{itemize}

\textit{A holistic design and optimization framework for low-carbon system hardware spanning fabrication processes, to architectures, to applications}, considering application-specific requirements and hardware provisioning is required. 

To address these outstanding challenges, we propose a system design and optimization framework to co-design embodied \textit{and} operational carbon---the overall life cycle carbon emissions---during the hardware development process. Our framework considers three key aspects of system design as inputs: \textit{embodied carbon parameters} specific to commercial fabrication of system hardware; \textit{operational use parameters}, such as non-operational idle time and use-phase carbon intensity measured in units of $\textrm{CO}_2$~emitted per kWh of energy; and \textit{characteristics of application use cases}, which system hardware is tailor-designed for. We formulate the design space of carbon-efficient systems as a multi-objective optimization problem, where we propose a new metric \cdp~that minimizes for total life cycle carbon and delay subject to power, area and Quality-of-Service (QoS) design constraints. A key contribution of our work is the consideration of the intricate interplay between embodied and operational carbon as the carbon bottleneck shifts based on the use and lifetime of computing systems. 

Our key contributions are:
\begin{enumerate}
    \item \textbf{Quantification of In-Production Systems \emph{using the\\tCDP metric}, which identifies significant opportunities for carbon efficiency improvements.} We take a data-driven approach to demonstrate that carbon aware metrics need to capture total life cycle carbon based on retrospective CPU and SoC analysis. Using in-production VR systems, we characterize the degree of dark silicon (i.e. under-utilized or unused circuitry) in terms of \textit{unused} embodied carbon (over 60\%) for real-world VR applications, highlighting significant carbon optimization potential for future system hardware (Section~\ref{sec:motivation}).
    \item \textbf{A Carbon-Efficient System Design Framework\\Co-optimizing Embodied and Operational Carbon:}. We propose a carbon efficiency optimization framework with a new metric, considering \textit{total life cycle carbon}.
    Using the proposed carbon efficiency metric \cdp, we simultaneously optimize for total life cycle carbon and PPA, achieving 9 and 49 times carbon efficiency improvement over designs optimized using CDP and CEP, respectively, for future XR system hardware (Section~\ref{sec:framework} and Section~\ref{sec:results-2}).
    \item \textbf{Case Studies in Designing Carbon Efficient Systems:} Our framework demonstrates the carbon efficiency improvement potential of application-specific hardware. Designing system hardware for AI versus XR workload use cases achieves up to 7.3 times better carbon efficiency improvement compared to system hardware tailor-designed for the entire AI and XR kernel suite (Section~\ref{sec:results-1}).
    Designing for specific hardware operational lifetime yield accelerators with 2.3 times better carbon efficiency (Section~\ref{sec:results-ar-vr-accelerators}). Leveraging hardware over-provisioning to optimize for carbon-efficient general-purpose hardware demonstrates up to 2 times embodied carbon savings and up to 21\% improvement in total life cycle carbon (Section~\ref{sec:results-existing-arvr-oculus}). Depending on operational lifetime and use of VR hardware, carbon-efficient optimal lifetimes vary between 5 years (20\% carbon savings) and 2 years (50\% carbon savings) (Section~\ref{sec:rebuttal-hardware-lifetime-replacement}). 
    For integrated circuits with stringent form factor requirements (as is the case for many XR applications), energy-intensive off-chip memory accesses, bandwidth limitations, advanced 3D-integration technologies achieve 1.1 to 7.86 times carbon efficiency improvements over 2D baseline (Section~\ref{sec:results-5}).
\end{enumerate}    
\section{Motivation}
\label{sec:motivation}

To understand the design and optimization space of carbon-efficient systems, we characterize and present the carbon characteristics of a variety of high-performance server-class CPUs and mobile SoCs (Section~\ref{sec:cpu-carbon-characterization}). Taking a step further, we expand the carbon characterization study to VR applications---an emerging application domain with significant carbon improvement potential (Section~\ref{sec:vr-carbon-characterization}). The data-driven insights motivate our framework for designing future carbon-efficient systems including systems for AI and XR.

\begin{figure}[t]
\centering
\includegraphics[width=1\columnwidth]{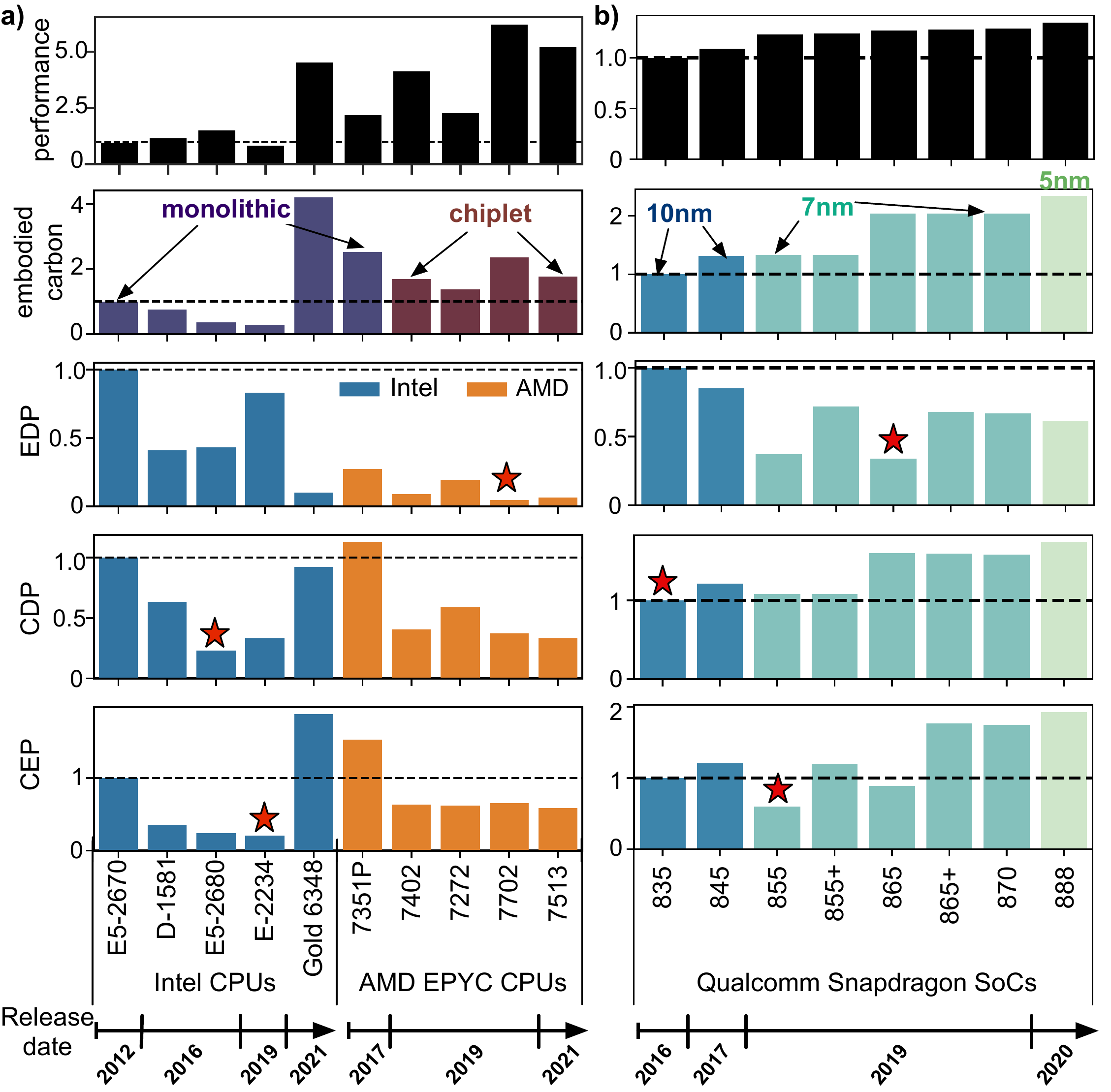}
\vspace{-0.7cm}
\caption{Carbon analysis of (a) Intel (blue) and AMD (orange) server-class CPUs released between 2012-2021. Top figure shows performance based on CPUMark (higher is better). (b) presents Qualcomm Snapdragon mobile SoCs released between 2016-2020, and performance based on CenturionMark (higher is better). Last four figures present embodied carbon, EDP and state-of-the-art carbon-aware metrics CDP and CEP. Stars indicate metric-optimal hardware. Results normalized to (a) Intel E5-2670 and (b) Snapdragon 835.}
\label{fig:CPUdatacenter}
\vspace{-1.1em}
\end{figure}

\subsection{Carbon-Aware Optimization Should Consider Total Life Cycle Carbon}
\label{sec:cpu-carbon-characterization}

Existing carbon metrics do not capture desired design optimization objectives sufficiently for life cycle emissions of system hardware. Figure~\ref{fig:CPUdatacenter}(a) presents the performance\footnote{CPU performance based on CPUMark benchmark suite~\citep{cpumark}. SoC performance based on CenturionMark benchmark suite~\citep{CenturionMark}. Operational energy estimate is $\textrm{E} =  \frac{\textrm{TDP}}{\textrm{Performance}}$, where TDP is the thermal design power.} and carbon analysis results over a variety of high-performance server-class CPUs (x-axis) released between 2012 and 2021\citep{wikichip, anadtechEPYC, anadtechIntel, techpowerup, cpuworld}, using energy-delay product (\textit{EDP}) and state-of-the-art carbon metrics~\citep{ACTGupta2022}---\textit{CDP} and \textit{CEP}.

Our analysis shows that the latest released CPUs and SoCs exhibit higher performance and lower operational energy. For Figure~\ref{fig:CPUdatacenter}(a) embodied carbon, we assume a fixed 80\% yield for monolithic CPUs (i.e. single die), and a higher yield for chiplet architectures (i.e. re-partitioning monolithic dies into multiple smaller dies) based on AMD's 0.59$\times$ cost reduction for chiplet designs versus the monolithic approach~\citep{AMD2021chiplet}. AMD chiplet CPUs exhibit embodied carbon benefits due to multiple smaller die areas with higher yield compared to AMD's EPYC 7351P, which is a larger monolithic die architecture. 

The estimation for fab carbon intensity assumes US grid for Intel CPUs~\citep{greencloudcomputing}, and Taiwan grid for AMD CPUs based on location. 
Given that EDP optimizes for operational energy and performance, the EDP-optimal CPU---AMD EPYC 7702---does not account for the CPU's embodied carbon or operational carbon intensity. The CDP-optimal CPU---Intel E5-2680---does not account for operational energy or carbon intensity of the datacenter's electrical grid. 

\begin{figure}[t!]
\centering
\includegraphics[width=0.95\columnwidth]{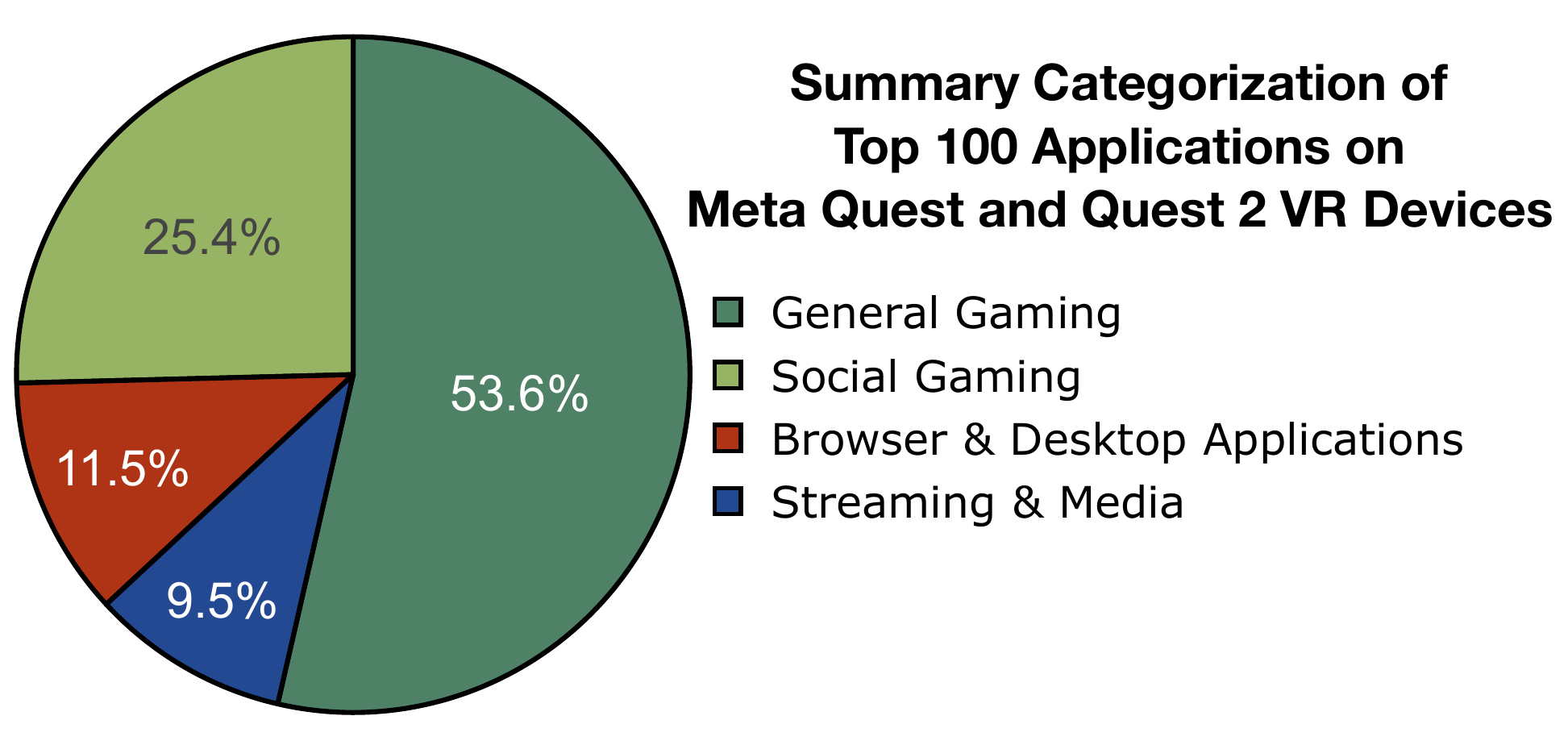}
\vspace{-0.4cm}
\caption{Summary categorization of the topmost 100 running applications on Meta Quest and Quest 2 VR devices in the wild. Gaming is the most dominant application category followed by social gaming. Top 10 applications cover $>$85\% of the total compute cycles for real-production VR headsets.}
\label{fig:oculusappschart}
\vspace{-1em}
\end{figure}

For mobile Qualcomm SoCs in Figure~\ref{fig:CPUdatacenter}(b), there is an increasing embodied carbon trend as process technology advances over the years. Similar to the CPU analysis, EDP-optimal SoC---Snapdragon 865---does not account for embodied carbon or operational carbon intensity. CDP-optimal SoC---Snapdragon 835---does not account for operational energy or the used electrical grid carbon intensity.

A challenge faced when we attempt to analyze the carbon efficiency of commodity CPUs and mobile SoCs is the lack of relationship between embodied and operational carbon. As illustrated in Figure~\ref{fig:CPUdatacenter}, metrics focused on embodied carbon only do not consider system hardware uses and therefore do not reflect carbon-efficient systems across the entire system product life cycle. This is because the metrics do not account for operational efficiency. For instance, while Intel E-2234 CPU is CEP-optimal (Figure~\ref{fig:CPUdatacenter}(a)), it is CDP sub-optimal because of the lower performance. Similarly for mobile SoCs, while Snapdragon 855 is CEP-optimal (Figure~\ref{fig:CPUdatacenter}(b)), it is CDP sub-optimal because of its higher embodied carbon. The existing metrics are insufficient to optimize for the hardware's energy efficiency without accounting for operational carbon.

Using the retrospective analysis, we show that 
\textit{EDP does not account for total life cycle carbon}, \textit{CEP is yet to capture the ratio between embodied and operational carbon of the system}, and \textit{CDP is yet to capture the operational energy and the carbon intensity of operational use}. Furthermore, carbon-aware metrics should capture the different design parameters of performance, operational energy efficiency, operational carbon and embodied carbon simultaneously. \textit{Designing system hardware with life-cycle emissions in mind leads to carbon-efficient systems}.

Given the rise of new application use cases, we expand our carbon characterization study to consumer-focused XR devices. With the rising use of XR systems, we analyze real-production VR devices (i.e. Meta Quest 2) to identify carbon reduction opportunities for carbon-efficient design decisions. Sections~\ref{sec:results-1} and~\ref{sec:results-2} demonstrate carbon efficiency improvement for future XR systems when \textit{total life cycle carbon} is a design parameter in the hardware optimization process. Designing environmentally-sustainable XR devices is particularly important as production volume of emerging XR systems is expected to multiply in the near future.

\begin{figure}[t!]
\centering
\includegraphics[width=\columnwidth]{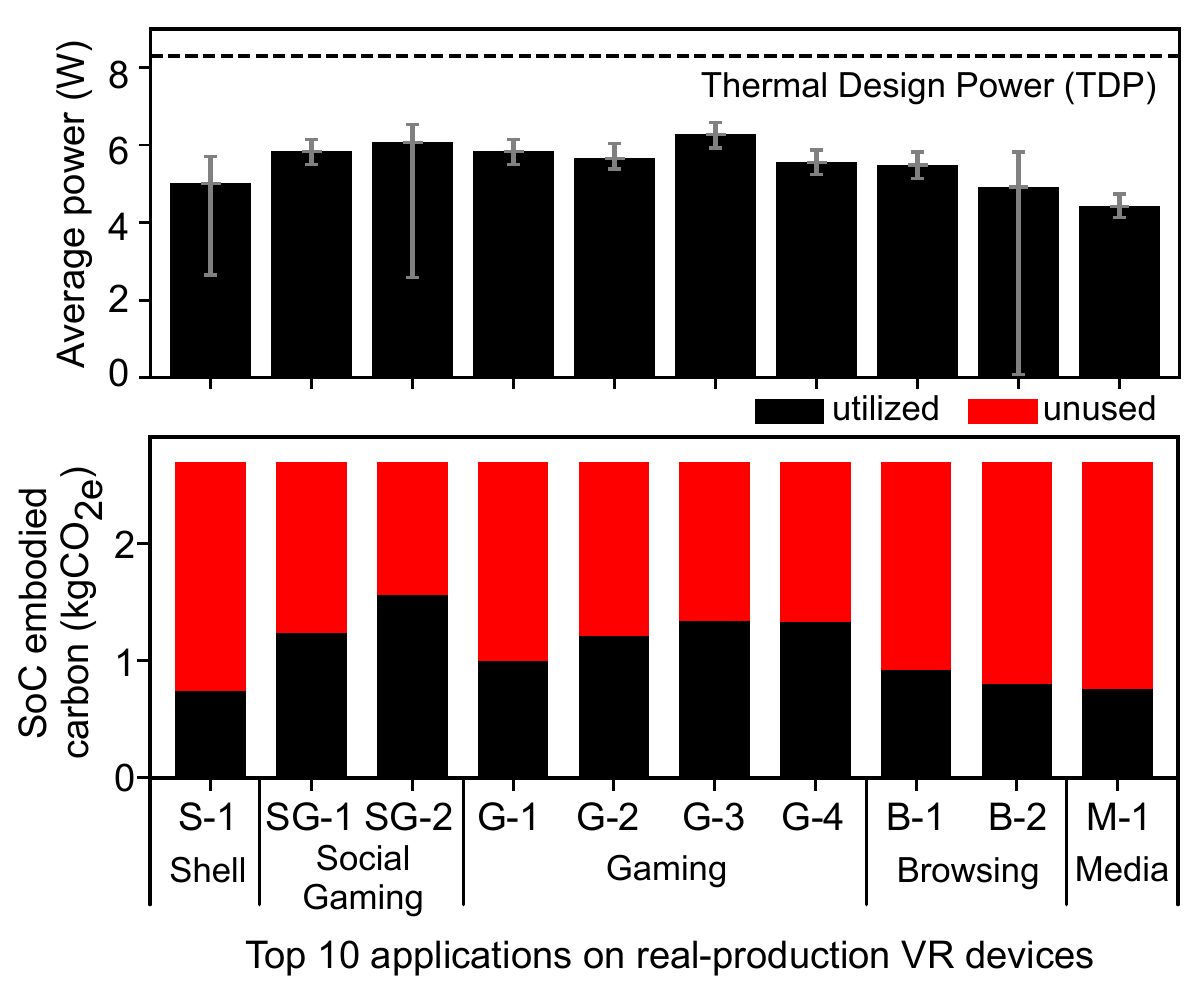}
\vspace{-0.85cm}
\caption{(Top) Average power consumed by the headset for each of the top 10 applications running on real-production Meta Quest 2 VR devices. Most applications utilize approximately 70\% of the device’s thermal design power budget (TDP). (Bottom) Embodied carbon footprint estimate of VR System-on-Chip (SoC). The split between unused (red) and utilized (black) embodied carbon is dependent on each application's hardware utilization. We observe embodied carbon reduction opportunities due to hardware over-provisioning.
}
\label{fig:oculus-apps-power-carbon}
\vspace{-1.1em}
\end{figure}

\begin{figure*}[t]
\includegraphics[width=\textwidth]{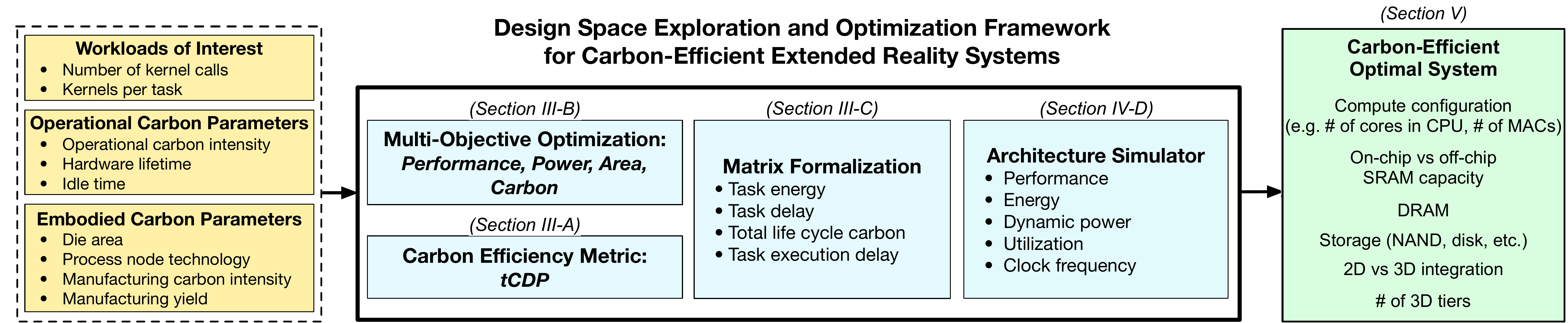}
\vspace{-0.6cm}
\caption{This work proposes a closed-loop design optimization framework for carbon-efficient XR systems.}
\label{fig:cordoba-block-diagram}
\vspace{-1em}
\end{figure*} 

\subsection{Hardware Over-Provisioning Presents Carbon Reduction Opportunities}
\label{sec:vr-carbon-characterization}

The optimization objectives for battery-powered systems are dominated by runtime demands---low latency and extremely low power. This has led to the carbon breakdown for state-of-the-art smartphones and wearables to pivot from operational to embodied carbon. The operational carbon to embodied carbon ratios for iPhone-3 and iPhone-14 released by \textcolor{black}{over} a decade apart have shifted from \textcolor{black}{49\% to 18\%}, respectively~\citep{iphone3Greport, iphone14report}. While high-level operational and embodied carbon breakdown is available for a wide collection of Apple products, to our knowledge, there has not been prior work that quantifies the potential carbon saving available for emerging applications.
Here, we take a data-driven approach to characterize VR applications in deployment, which we expect to see increasing relevance with the rise of VR devices. To harvest the at-scale impact over billions of VR systems in the years forward, it is crucial to understand their carbon optimization potential.

To design low-carbon system hardware for real-world VR applications, we analyze the top 100 applications running on real-production VR headsets based on the data collected from deployed VR devices running in the wild. We categorize the applications into four primary categories: general gaming (G), social gaming (SG), browser \& virtual desktop (B), and streaming \& media (M) as shown in Figure~\ref{fig:oculusappschart}. \textit{The top 10 applications cover over $85\%$ of the overall compute cycle} on all of the in-the-field VR devices.

Based on the measurement data, Figure~\ref{fig:oculus-apps-power-carbon} (top) presents the average power consumption of the top 10 VR applications with respect to the Thermal Design Power (TDP) of the VR headset. The error bars represent the p5 and p95 power consumption of the VR devices under study. Operational carbon follows same trend as operational power. Additionally, Figure~\ref{fig:oculus-apps-power-carbon} presents the embodied (black and red bars) carbon for the top 10 VR applications. The carbon characterization is based on the assumption that a real-production VR headset is used for one hour daily for a total device lifetime of three years. The embodied carbon footprint considers the CPU and GPU on the VR headset. We take a step further to breakdown the overall embodied carbon into \textit{utilized} and \textit{unused} embodied carbon based on hardware utilization, where utilization is defined as the active time of the hardware over the total application runtime. 

Our goal is to reduce the embodied carbon footprint of computing while maintaining a competitive level of performance and operational efficiency. While it is important to account for peak performance, in most usage scenarios, users do not utilize the hardware components sufficiently. Therefore, there is \textit{unused} hardware real estate on devices, which results in embodied carbon inefficiencies. Hardware under-utilization is also observed at the datacenter scale~\citep{CaseforEnergyProportionalComputing2007,acun:asplos2023,wu:mlsys2022}.
Last but not least, current computations of embodied carbon do not account for hardware utilization of the SoC. In many cases, especially in consumer electronics, devices are over-provisioned with the majority of hardware sitting idle or in the sleep mode.
 
When the operational time includes both idle and non-idle runtime, possible embodied carbon reduction opportunities may be overlooked. With the ever-increasing need to design environmentally-sustainable computing systems, it is necessary to account for any under-utilized hardware real estate in future systems.
\section{Carbon-efficient Optimization Framework}
\label{sec:framework}

We propose a design space exploration and optimization framework to design carbon-efficient computing systems. It enables the consideration of the dynamic interplay between the operational \textit{and} embodied carbon at the design phase. This unique, new feature allows architects and hardware designers to produce systems with improved carbon efficiency.
Figure~\ref{fig:cordoba-block-diagram} provides an overview for the proposed closed-loop carbon-aware design and optimization framework. 

The framework takes in three input categories: (1) Embodied carbon parameters for the hardware components and the manufacturing conditions; (2) Operational use parameters including use-phase carbon intensity and system idle time; (3) Workloads of interest including the number of kernel calls and kernels per task to enable multi-workload carbon-efficient optimization. 

We formulate the design space of carbon-efficient systems as a multi-objective optimization problem. The input parameters are encoded into matrices and we solve the optimization objective subject to the design constraints by minimizing \cdp---\textit{total} Carbon Delay Product, where C is the life cycle carbon of the system.

Section~\ref{sec:metric} introduces the new carbon efficiency metric: \cdp and Section~\ref{sec:mathematical-formulation} details the mathematical formulation of the carbon efficiency optimization problem. Lastly, Section~\ref{sec:matrix-formulation} describes the matrix formalization of the carbon efficiency design optimization parameters.  

\begin{table*}[t]
\centering
\caption{Defining $\beta$ in Optimization Framework.}
\label{tab:beta}
\vspace{-0.9em}
\scalebox{0.92}{
\begin{tabular}{|c|c|c|}
\hline
\textbf{$\beta$ Parameter} & \textbf{Carbon-efficiency Equation} &\textbf{Design Use-case} \\ 
\hline \hline
$\beta\xrightarrow{}0$ & $\textrm{C}_{\textrm{operational}} \cdot \textrm{D}$ & clean fab \& operational carbon dominant system \\  \hline 
$\beta\xrightarrow{}\infty$ & $\textrm{C}_{\textrm{embodied}} \cdot \textrm{D}$ & 100\% renewable energy-grid \\ \hline 
$0<\beta<1$ & $(\textrm{C}_{\textrm{operational}} + \beta \cdot \textrm{C}_{\textrm{embodied}}) \cdot \textrm{D}$ & operational carbon dominance range \\ \hline
$1<\beta<\infty$ & $(\textrm{C}_{\textrm{operational}} + \beta \cdot \textrm{C}_{\textrm{embodied}}) \cdot \textrm{D}$ & embodied carbon dominance range \\ \hline 
$\beta = 1$ & $(\textrm{C}_{\textrm{operational}} + \textrm{C}_{\textrm{embodied}}) \cdot \textrm{D}$ & embodied and operational carbon are in \co~units and relation known at design phase \\ \hline
\end{tabular}}
\vspace{-0.8em}
\end{table*}

\subsection{\cdp: A Figure-of-Merit to Guide Carbon-Aware System Optimization}
\label{sec:metric}

To effectively evaluate the carbon efficiency of new system hardware designs using our optimization framework, we propose a primary carbon efficiency figure-of-merit where: 
\begin{equation*}
    \textrm{\cdp} = \textrm{C}_{\textrm{total}} \cdot \textrm{Task execution delay}
\end{equation*}
\cdp~optimizes for the total life cycle carbon footprint of the system ($\textrm{C}_{\textrm{total}}$), i.e. operational \textit{and} embodied, and simultaneously trade-off the conventional hardware optimization approaches of performance-power-area (PPA) and energy efficiency. 
\cdp~is a combined optimization metric that gives additional information beyond the isolated metrics of carbon and delay. Specifically, \cdp~quantifies how efficiently we are using the carbon of the systems we design. Lower \cdp~indicates better carbon-efficient systems.

\subsection{Optimization Framework}
\label{sec:mathematical-formulation}

To optimize for low-carbon computing systems while achieving operational efficiency, the framework minimizes for embodied carbon, operational carbon, and delay. We describe the objectives of our optimization problem as follows:
\begin{equation*}
\begin{aligned}
& F_1(x) = \textrm{C}_{\textrm{operational}}(x) \cdot \textrm{D}(x) \\
& F_2(x) = \textrm{C}_{\textrm{embodied}}(x) \cdot \textrm{D}(x)  \\
& \underset{x}{\text{minimize}}~~f_0(x) = F_1(x) + F_2(x)
\end{aligned}
\end{equation*}

where $x$ represents the hardware computing system, $\textrm{C}_{\textrm{operational}}$(x) for operational carbon, $\textrm{C}_{\textrm{embodied}}$(x) for embodied carbon, and $D(x)$ for the task execution delay of the system. Our goal is to minimize for operational carbon, embodied carbon and small task execution delay (i.e. small $F_1(x)$ and small $F_2(x)$). To address the challenge of not knowing the trade-offs due to carbon accounting uncertainties, we scalarize the multi-objective optimization problem into $\lambda_1 F_1(x) + \lambda_2 F_2(x)$ and minimize the weighted scalar sum objective. Without loss of generality, we can take $\lambda_1 = 1$ and $\lambda_2 = \beta > 0$ when forming the associated scalarized problem to optimize for carbon-efficient computing systems:

\fixme{\begin{equation*}
\begin{aligned}
& \underset{x}{\text{minimize}} 
& & (\textrm{C}_{\textrm{operational}}(x) + \beta \cdot \textrm{C}_{\textrm{embodied}}(x)) \cdot \textrm{D}(x)  \\
& \text{subject to}
& & \textrm{Area}_i(x) \leq a_i,\tab[1.18cm] i = 1,\ldots, I\\
& 
& & \textrm{QoS}_j(x) \geq \fixme{q}_j,\tab[1.17cm] j = 1,\ldots, J\\
& 
& & \textrm{Power}_l(x) \leq p_l,\tab l = 1,\ldots, L\\
\end{aligned}
\end{equation*}}
Here $x$ represents the whole system consisting different logic and memory components. Parameters a$_i$, q$_j$, and p$_l$ are the area, quality of service (QoS), and power optimization constraints of the different components of the system, respectively.

$\beta$ is equal to 1 when $\textrm{C}_{\textrm{operational}}$ and $\textrm{C}_{\textrm{embodied}}$ of a system are both in units of \co~and their relative ratio is known to compute the total carbon of the system $x$. In this case, $F_1(x) + \beta \cdot F_2(x) = (\textrm{C}_{\textrm{operational}}(x) + \textrm{C}_{\textrm{embodied}}(x)) \cdot \textrm{D}(x)$ exactly (refer \cdp~Subsection~\ref{sec:metric}). However, there are multiple reasons why the relation between embodied carbon and operational carbon of the system could be unknown. For instance, due to uncertainty in the quantification of carbon footprint data; lack of knowledge on conditions and period of operational use; or when the parameters to compute \co~are challenging to identify their relative scale factors.
Using $\beta$ we can sweep the pareto-optimal curve of $F_1(x)$ versus $F_2(x)$ based on the primary design constraints and trade-offs between the different design points. Therefore, when the relative scaling between embodied and operational carbon is challenging to quantify, we mathematically know the true carbon-efficient optimal point is somewhere on the pareto-optimal front. Table~\ref{tab:beta} defines the sweep ranges of $\beta$. 

Below we present a mathematical example of the optimization framework for real-production VR headsets:
\begin{equation*}
\begin{aligned}
& \underset{x}{\text{minimize}} 
& & (\textrm{C}_{\textrm{operational}}(x) + 1 \cdot \textrm{C}_{\textrm{embodied}}(x)) \cdot \textrm{D}(x)  \\
& \text{subject to}
& & \textrm{Area}_{\textrm{SoC}}(x) \leq 2.25\textrm{cm}^{2}, \\
&
& & \textrm{Area}_{\textrm{CPU}}(x) \leq 0.45\textrm{cm}^{2},\\
& 
& & \textrm{QoS}(x) \geq 60 \textrm{FPS},\\
& 
& & \textrm{Power}_{\textrm{SoC}}(x) \leq 8.3 \textrm{W}\\
\end{aligned}
\end{equation*}
Where $x$ represents the VR headset system, the area constraints are based on data presented in Methodology Section~\ref{sec:methodology-carbon-measurement} and Table~\ref{tab:app_areas}, the QoS constraint is a target frame rate to ensure quality of user-experience~\citep{ILLIXR2020}, and the power constraint is 8.3W which is the TDP previously shown in Figure~\ref{fig:oculus-apps-power-carbon}. The corresponding carbon efficiency optimal results are discussed in Section~\ref{sec:results-existing-arvr-oculus}.

\begin{table}[t!]
\centering
\caption{Matrix formalization parameters.}
\label{tab:parameters}
\vspace{-0.9em}
\scalebox{0.89}{
\begin{tabular}{|c|c|c|}
\hline
\textbf{Parameter}   & \textbf{Description}  \\
\hline \hline
k & Kernels \\
N & Number of kernel calls \\
T & Task defined by number of kernel calls \\
D & Task delay \\
$x$ & Hardware target system \\
U & Hardware utilization ratio \\
E & Operational energy consumption \\
P$_{\textrm{leakage}}$ & Leakage power \\
P$_{\textrm{dynamic}}$ & Dynamic power \\
P$_{\textrm{total}}$ & Total power = P$_{\textrm{leakage}}$ + P$_{\textrm{dynamic}}$ \\
$f_{\textrm{clk}}$ & Clock frequency \\
CI$_{\textrm{use}}$ & Use-phase carbon intensity \\
\textrm{CI}$_{\textrm{fab}}$ & Fab carbon intensity \\
\textrm{EPA} & Fab energy per area \\
\textrm{MPA} & Materials per area \\
\textrm{GPA} & Direct gases emitted per area \\
\textrm{Y} & Fab yield \\
\textrm{A} & Area of \fixme{components} in system $x$ \\
C$_{\textrm{operational}}$ & Operational carbon \\
C$_{\textrm{embodied, }x_i}$ & Embodied carbon per component in system $x$\\
C$_{\textrm{embodied, overall}}$ & Overall embodied carbon of a system \\
LT & Hardware lifetime \\
D$_{\textrm{idle}}$ & Idle time throughout the system's lifetime \\
C$_{\textrm{embodied}}$ & Amortized embodied carbon \\
\cdp & Carbon efficiency metric \\
\hline 
\end{tabular}}
\vspace{-0.8em}
\end{table}

\subsection{Matrix Formalization}
\label{sec:matrix-formulation}
Next, we formulate the key inputs and parameters of the design optimization in a matrix format. The matrix formalization enables designers to tune design knobs and identify hardware configurations that maximize carbon efficiency. 

Table~\ref{tab:parameters} presents the parameters of the matrix formalization. We define a task T as a set of k kernels. Where each task could be one kernel or more, depending on the number of kernel calls per task ($\textrm{N}_{\textrm{T}, \textrm{k}}$). For example, a zero value would indicate that a kernel k is not part of the task T. This formulation allows us to optimize for multiple computation kernels and application tasks of interest, while meeting the latency performance requirements and hardware design constraints of the system. 

Below we present a detailed breakdown of the four main computations in the matrix formalization:
\subsubsection{\textbf{Task energy}}

First, we compute the energy consumption per task ($\textrm{E}_{\textrm{T}}$). Each kernel in a task consumes energy based on clock frequency, $f_{\textrm{clk}}$, hardware-dependent leakage power, $\textrm{P}_{\textrm{leakage}}$, and dynamic power, $\textrm{P}_{\textrm{dynamic}}$. To get the energy consumption for all tasks, we multiply number of kernel calls per task ($\textrm{N}_{\textrm{T}, \textrm{k}}$) \fixme{with the energy consumption} ($\frac{\textrm{P}_{\textrm{leakage}}}{f_{\textrm{clk}}} + \frac{\textrm{P}_{\textrm{dynamic}}}{f_{\textrm{clk}}}$) per kernel. We get task execution energy vector \textbf{E}:

\begin{gather*}
 \begin{bmatrix} 
  \textrm{N}_{\textrm{T}_{1}, \textrm{k}_{1}}&..&\textrm{N}_{\textrm{T}_{1}, \textrm{k}_{n}}\\
  :&:&: \\
  \textrm{N}_{\textrm{T}_{m}, k_{1}}&..&\textrm{N}_{\textrm{T}_{m}, \textrm{k}_{n}} \\
 \end{bmatrix} \times 
 \fixme{(}
 \begin{bmatrix}
     \frac{\textrm{P}_{\textrm{leakage}}}{f_{\textrm{clk, 1}}} \\ 
     : \\
     \frac{\textrm{P}_{\textrm{leakage}}}{f_{\textrm{clk, n}}} \\
 \end{bmatrix} +
 \begin{bmatrix}
     \frac{\textrm{P}_{\textrm{dynamic}}}{f_{\textrm{clk, 1}}} \\ 
     : \\
     \frac{\textrm{P}_{\textrm{dynamic}}}{f_{\textrm{clk, n}}} \\
 \end{bmatrix} 
 \fixme{)}
 = 
 \begin{bmatrix}
    \textrm{E}_{\textrm{T}_{1}} \\
    : \\
    \textrm{E}_{\textrm{T}_{m}} \\
 \end{bmatrix}
\end{gather*}

\subsubsection{\textbf{Task delay}}

Second, to get the task delay per task ($\textrm{D}_{\textrm{T}}$), we multiply the number of kernel calls per task ($\textrm{N}_{\textrm{T}, \textrm{k}}$) and the kernel execution time ($\textrm{D}_{\textrm{k}}$) on its respective hardware system. We get the task delay vector \textbf{D}:
\begin{gather*}
 \begin{bmatrix} 
  \textrm{N}_{\textrm{T}_{1}, \textrm{k}_{1}}&..&\textrm{N}_{\textrm{T}_{1}, \textrm{k}_{n}}\\
  :&:&: \\
  \textrm{N}_{\textrm{T}_{m}, \textrm{k}_{1}}&..&\textrm{N}_{\textrm{T}_{m}, \textrm{k}_{n}} \\
 \end{bmatrix} \times 
 \begin{bmatrix}
     \textrm{D}_{\textrm{k}_{1}} \\
    : \\
    \textrm{D}_{\textrm{k}_{n}} \\
 \end{bmatrix} 
 = 
 \begin{bmatrix}
    \textrm{D}_{\textrm{T}_{1}} \\
    : \\
    \textrm{D}_{\textrm{T}_{m}} \\
 \end{bmatrix}
\end{gather*}

\subsubsection{\textbf{Total life cycle carbon}}

We define the total life cycle carbon of executing a task on a hardware target to be the sum of both operational \textit{and} embodied carbon footprint, where $\textrm{C}_{\textrm{total}} = \textrm{C}_{\textrm{operational}} + \textrm{C}_{\textrm{embodied}}$. Below we detail how we compute operational and embodied carbon:

\begin{itemize}
    \item \underline{\textbf{Operational carbon}}

To compute the operational carbon ($\textrm{C}_{\textrm{operational}}$) due to the energy consumption of the task operating on hardware, we multiply the use-phase carbon intensity (CI$_{\textrm{use}}$) with the $L1$ norm of the task energy vector ($||\textbf{E}||_1$): 

\begin{gather*}
 \textrm{CI}_{\textrm{use}} \times
 \begin{Vmatrix}
  \textbf{E}
 \end{Vmatrix}_{1}
 = \textrm{C}_{\textrm{operational}}
\end{gather*}

The above equation is analogous to the scalar operational carbon equation presented in ACT\footnote{Architectural Carbon Modeling Tool used to quantify carbon footprint.}~\citep{ACTGupta2022}. \\

    \item \underline{\textbf{Embodied carbon}}

To compute the overall embodied carbon of a chip, we use ACT's embodied carbon equation as follows~\citep{ACTGupta2022}: 
\begin{equation*}
    \textrm{C}_{\textrm{embodied, }x_i} = (\textrm{CI}_{\textrm{fab}} \times \textrm{EPA} + \textrm{MPA} +  \textrm{GPA}) \times \frac{\textrm{A}}{\textrm{Y}} 
\end{equation*}

where $\textrm{CI}_{\textrm{fab}}$ stands for carbon intensity of the fab's electrical grid, EPA is fab energy per die area, MPA is carbon footprint of procured materials used in manufacturing per area, GPA is direct fab gas emissions, A is die area, and Y is fab yield. These parameters are dependent on the fab vendor and the process technology node. 

To compute the overall embodied carbon of hardware target $x$, we multiply the embodied carbon hardware target vector (\textrm{C}$_{\textrm{embodied, $x$}}$) with a binary vector, indicating online (1) and offline (0) components in the hardware target system. The different components in the embodied carbon hardware target vector indicate different compute and memory configurations, such as number of CPU cores, number of Multiply-Accumulate (MAC) Arrays, GPUs, DRAM, on-chip versus off-chip SRAM, and DSPs. 
\begin{gather*}
    \begin{bmatrix}
        \textrm{C}_{\textrm{embodied, $x_{1}$}} & .. & \textrm{C}_{\textrm{embodied, $x_{i}$}} 
    \end{bmatrix} \times 
    \begin{bmatrix}
        0 \\
        : \\
        1
    \end{bmatrix} = \textrm{C}_{\textrm{embodied, overall}}
\end{gather*}

where $\textrm{C}_{\textrm{embodied, overall}}$ is the overall embodied carbon of the system throughout its lifetime. This formulation provides hardware provisioning as a design knob in our optimization framework. Therefore, designers can optimize for carbon efficiency presented by \textit{unused} embodied carbon reduction opportunities. In Section~\ref{sec:results-existing-arvr-oculus}, we showcase carbon efficiency improvements in real-production VR systems due to hardware provisioning design optimization.

Next, we amortize embodied carbon over the execution time and not the lifetime of the system in years. This is to ensure we do not underestimate and amortize the embodied carbon over idle time when the system is not in use. We define $\textrm{D}_{\textrm{idle}}$ as the idle time throughout the system's lifetime (LT). Where $\textrm{LT} - \textrm{D}_{\textrm{idle}}$ is the system's operational lifetime. We take the $L1$ norm of the task delay vector \textbf{D} to get the total task delay ($||\textbf{D}||_{1}$).

Amortized embodied carbon ($\textrm{C}_{\textrm{embodied}}$) is as follows:
\begin{equation*}
    \textrm{C}_{\textrm{embodied}} = \textrm{C}_{\textrm{embodied, overall}} \times \frac{||\textbf{D}||_{1}}{\textrm{LT} - \textrm{D}_{\textrm{idle}}}
\end{equation*}
\end{itemize}

\subsubsection{\textbf{Task execution delay}}

To optimize for carbon-efficient systems, i.e. high performing and low-carbon systems, we minimize for the reciprocal of performance or the task execution delay. Task execution delay can be computed by taking the $L1$ norm of the task delay vector (\textbf{D}) to get the total task delay ($||\textbf{D}||_{1}$). It can also be computed as the reciprocal of another performance measurement relevant for the system optimization space, e.g. CPUMark, SPEC scores, frames-per-second.

\section{Experimental Methodology}
\label{sec:methodology}

\begin{figure}[t!]
\centering
\includegraphics[width=0.95\columnwidth]{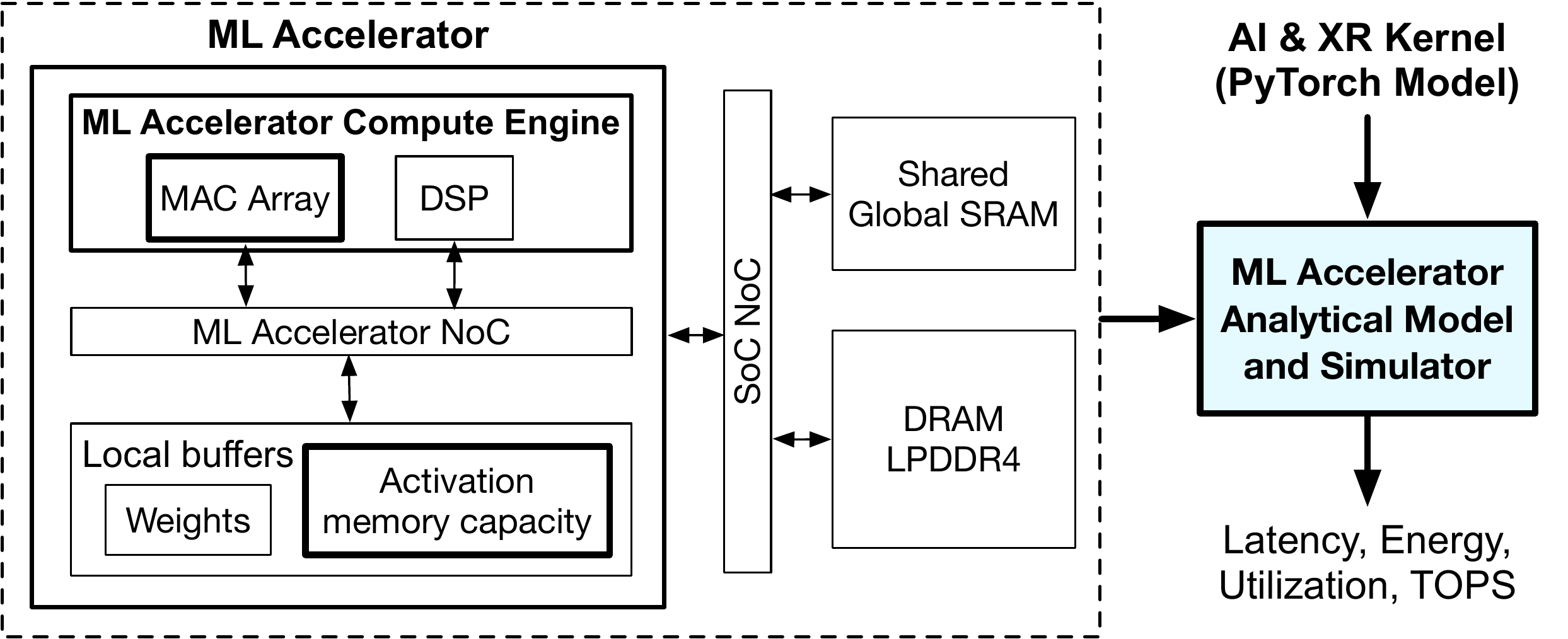}
\vspace{-0.3cm}
\caption{Hardware accelerator simulator used in optimization framework.}
\label{fig:inhouse-ml-simulator-block-diagram}
\vspace{-1.1em}
\end{figure}

\subsection{Workloads}
\label{sec:methodology-workloads}

\begin{table}[t!]
\centering
\caption{Summary of the AI and XR workloads profiled.}
\label{tab:workloads}
\vspace{-0.9em}
\scalebox{0.87}{
\begin{tabular}{|l|l|c|}
\hline
\textbf{Task} & \textbf{Computation Kernel}  & \textbf{Category} \\ 
\hline \hline
Object classification  & \begin{tabular}[c]{@{}l@{}}ResNet-18 (\textbf{RN-18})~\citep{resnet} \\ ResNet-50 (\textbf{RN-50})~\citep{resnet} \\ 
ResNet-152 (\textbf{RN-152})~\citep{resnet} \\
GoogleNet (\textbf{GN})~\citep{GoogleLeNet}\end{tabular} & AI \\ \hline
Object detection & 
MobileNet-V2 (\textbf{MN2})~\citep{mn2} & AI \\ \hline
Eye tracking & SegNet (\textbf{ET})~\citep{segnet} & XR \\ \hline
Depth estimation  & 3D Aggregation (\textbf{3D-Agg})~\citep{3dagg} & XR \\ \hline
\begin{tabular}[c]{@{}l@{}} Depth estimation \\ for augmented calls\end{tabular} & High-Res Net (\textbf{HRN})~\citep{hrn} & XR \\ \hline
Emotion detection & EmoFAN (\textbf{E-FAN})~\citep{emofan} & XR \\ \hline
Hand tracking & Joint Location Predictor (\textbf{JLP})~\citep{ht} & XR \\ \hline
Image denoising  & \textbf{UNet}~\citep{unet} + Feature-Align (\textbf{DN})~\citep{FeatureAlign} & XR \\ \hline
Super-resolution & Superres (\textbf{SR})~\citep{sr} & XR \\ \hline 
\end{tabular}}
\vspace{-0.7em}
\end{table}
\begin{table}[t!]
\centering
\caption{Summary of design space exploration kernel clusters.}
\label{tab:kernel-clusters}
\vspace{-0.9em}
\scalebox{0.9}{
\begin{tabular}{|c|c|}
\hline
\textbf{Cluster name} & \textbf{Kernels}  \\ 
\hline \hline
10 XR-dominant & \begin{tabular}[c]{@{}c@{}}
3D-Agg; ET; JLP; HRN; UNet; E-FAN; DN; \\
SR (256$\times$256); SR (512$\times$512); SR (1024$\times$1024) 
\end{tabular} \\
\hline
10 AI-dominant &  \begin{tabular}[c]{@{}c@{}}
RN-18; RN-50; RN-152; GN; MN2;\\
3D-Agg; ET; UNet; JLP; HRN; 
\end{tabular} \\
\hline
5 XR & 3D-Agg; HRN; DN; SR (512$\times$512); SR (1024$\times$1024)\\
\hline
5 AI & RN-18; RN-50; RN-152; GN; MN2\\
\hline
\end{tabular}}
\vspace{-0.7em}
\end{table}
We characterize a variety of AI and XR workloads, shown in Table~\ref{tab:workloads}, using the hardware performance profiling tool specified in Section~\ref{sec:methodology-accelerator-profiler} below. 
For our design space exploration, we cluster the AI and XR workloads into five clusters; 10 XR-dominant, 10 AI-dominant, 5 XR, and 5 AI as specified in Table~\ref{tab:kernel-clusters}.

\begin{figure*}[t!]
\centering
\includegraphics[width=1.01\textwidth]{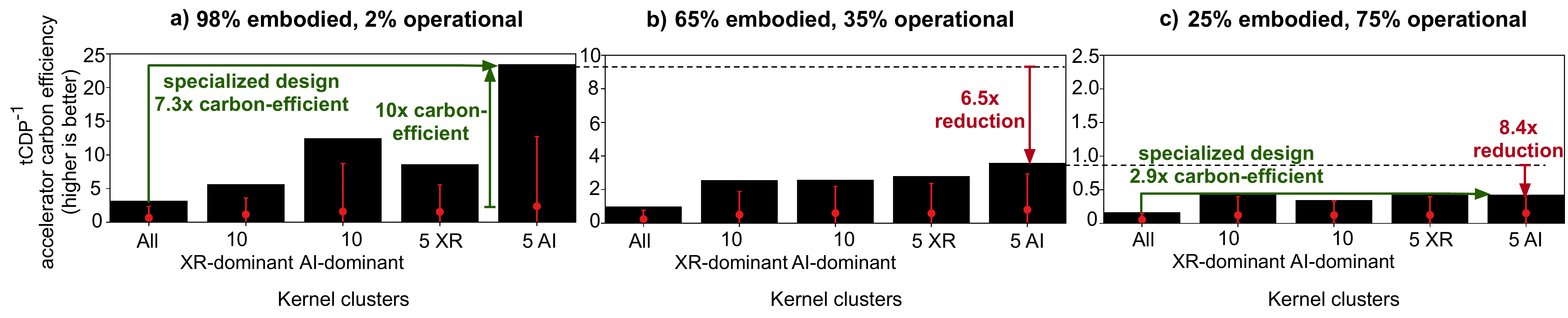}
\vspace{-0.8cm}
\caption{Carbon efficiency trends of our design space exploration normalized to the \underline{All} kernel cluster corresponding to 65\% embodied to total life cycle carbon in (b). Red dots represent the average carbon efficiency of the accelerators produced in the design space exploration. Error bars indicate the p5 and p95 carbon-efficient accelerators. Green arrows indicate higher carbon efficiency trends, while red arrows indicate lower carbon efficiency.}
\label{fig:summary_plot}
\vspace{-1em}
\end{figure*}

\subsection{Carbon Measurement}
\label{sec:methodology-carbon-measurement}
To compute the embodied carbon footprint, we use ACT which is an architectural carbon footprint modeling tool based on industry environmental reports and detailed fab characterization~\citep{ACTGupta2022}. We update ACT to include the most recent fab characterization data~\citep{imec2022logic} and incorporated more die placement and yield models~\citep{Vries2005dieplacementmodel, murphyYield}.

We compute the embodied carbon of the SoC used for real-production VR headsets based on the following assumptions. The Qualcomm Snapdragon SoC is manufactured in 7nm process technology node and has an Octa-core CPU. We approximate 20\% of the SoC is devoted to each of the CPU, based on the annotated floorplan die photo of a Qualcomm Snapdragon 845 by TechInsights~\citep{anandtech2018snapdragon, techinsights2018S9teardown}. We approximate the gold cores are $\frac{2}{3}$ of the CPU's total area and silver cores are the remaining $\frac{1}{3}$. We assume a fixed 85\% die yield and assume coal-grid carbon intensity for the manufacturing electricity. The estimates are summarized in Table~\ref{tab:app_areas}. 

\begin{table}[ht!]
\vspace{-0.4em}
\centering
\caption{Area and embodied carbon estimations of CPU gold and silver cores for real-production VR SoCs.}
\label{tab:app_areas}
\vspace{-0.9em}
\scalebox{0.9}{
\begin{tabular}{|l|c|c|}
\hline
\textbf{Parameter}   & \textbf{VR}  \\
\hline \hline
Total die area (cm$^2$) & 2.25  \\
CPU (cm$^2$) & 0.45  \\
CPU gold (cm$^2$) & 0.3 \\
CPU silver (cm$^2$) & 0.15\\
CPU gold embodied carbon (gCO$_{2}$e) & 895.89 \\
CPU silver embodied carbon (gCO$_{2}$e) & 447.94\\
\hline 
\end{tabular}}
\vspace{-1.3em}
\end{table}

\subsection{Production-level VR System Hardware}
\label{sec:methodology-oculus}
We use the Android Debug Bridge (adb) software to connect to the unlocked VR device and profile applications in real-time. Using Simpleperf~\citep{simplerperf}, a native CPU profiling tool for Android, we collect the number of instructions and clock cycles for each core configuration to compute Instructions-per-cycle (IPC) per application. We also measure the frame rate (FPS), temperature, and GPU utilization\% every 1s. Additionally, we use Perfetto UI~\citep{perfetto}, an open-source system-wide profiling and application tracing software for Android, to profile the thread-level parallelism (TLP) of the top 10 applications running on VR. 

\subsection{Performance and Power Simulation for Hardware Design}
\label{sec:methodology-accelerator-profiler}
We use an accelerator simulator based on a scaled-up version of Simbal et al's work~\citep{accelerator2022Meta}. The simulator takes a neural network model as input, extracts the operators, and outputs Tera-Operations-per-Second (TOPS) performance, latency, utilization and energy consumption for specified hardware architecture configurations (Figure~\ref{fig:inhouse-ml-simulator-block-diagram}). We evaluate and present the performance and energy results for four distinct hardware accelerators based on production hardware targets. 

\section{Evaluation Results and Analysis}
\label{sec:results}
\subsection{Result Overview for the Carbon-Efficient System Design Space Exploration}
\label{sec:results-1}
 
Figure~\ref{fig:summary_plot} presents the carbon efficiency results (y-axis) of the hardware design space exploration. The x-axis represents the five application workload clusters\footnote{Table~\ref{tab:workloads} summarizes the AI and XR kernels in each cluster.} system hardware is tailor-designed for using our framework. The bars represent the carbon efficiency of the optimal hardware accelerator configurations using \cdp. Whereas the dots represent the average carbon efficiency results of the hardware accelerator across the hardware design space of the 121 Multiply-Accumulate (MAC) arrays and on-chip SRAM capacity configurations. The error bars show the p5 and p95 carbon-efficient design points. 
To demonstrate our carbon-efficient design optimization framework capturing the embodied and operational carbon trade-offs, we design for different workload capacities and operational lifetimes corresponding to 98\%, 65\%, and 25\% embodied carbon to the total life cycle carbon. 

Assuming same hardware lifetime and utilization, when embodied carbon is the dominating source of total life cycle carbon, specializing system hardware for applications of similar characteristics can lead to  hardware with higher carbon efficiency. For example, the most carbon efficient hardware accelerator tailor-designed for the \underline{5 AI} kernel cluster is \textit{7.3 times more carbon-efficient} than the most carbon-efficient accelerator designed for \underline{All} (Figure~\ref{fig:summary_plot}(a)). This also applies when operational carbon is the dominating source of the total life cycle carbon. For example, the most carbon-efficient accelerator designed for \underline{5 AI} is 2.9 times more carbon-efficient than the the most carbon-efficient accelerator designed for \underline{All} (Figure~\ref{fig:summary_plot}(c)). 

Furthermore, the room for carbon efficiency improvement is significant. Carbon efficiency quantified by \cdp~can be 10 times higher for the most carbon efficient hardware accelerator configuration than the average accelerator (Figure~\ref{fig:summary_plot}(a) \underline{5 AI} kernel cluster).

When designing system hardware for longevity in mind with a lower embodied to life cycle carbon ratio (Figure~\ref{fig:summary_plot}(a) to (c)), the drastic carbon improvement potential is diminished. For example, for the \underline{5 AI} kernel cluster, the carbon efficiency improvement is reduced from 6.5 to 8.4 times when the embodied to total life cycle carbon ratios move from 98\% to 25\%. This demonstrates the importance of considering total life cycle carbon into the hardware design process---both embodied \textit{and} operational carbon. Thus, we achieve 1.29 times carbon efficiency improvement over the two distinct product use longevity scenarios.

\subsection{Carbon Efficiency of \cdp~over Existing Metrics}
\label{sec:results-2}

\begin{figure}[t!]
\centering
\includegraphics[width=0.9\columnwidth]{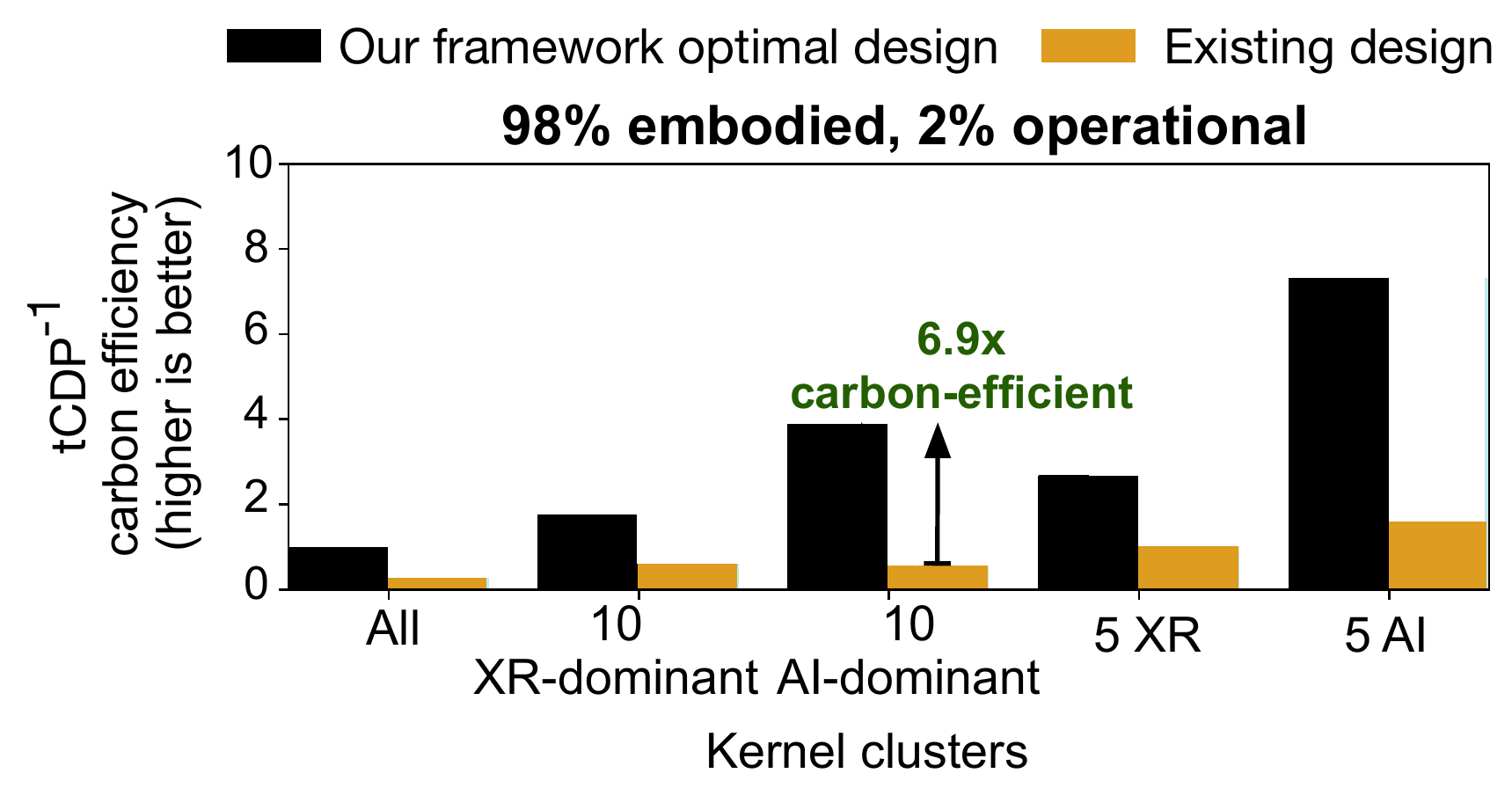}
\vspace{-0.4cm}
\caption{Carbon efficiency benefits of optimizing for total life cycle carbon versus the carbon efficiency of optimizing for a carbon oblivious metric---\textit{EDP}. Results are normalized to the \underline{All} kernel clusters.}
\label{fig:comparison_plot}
\vspace{-1.1em}
\end{figure}

\begin{figure}[t!]
\centering
\includegraphics[width=0.9\columnwidth]{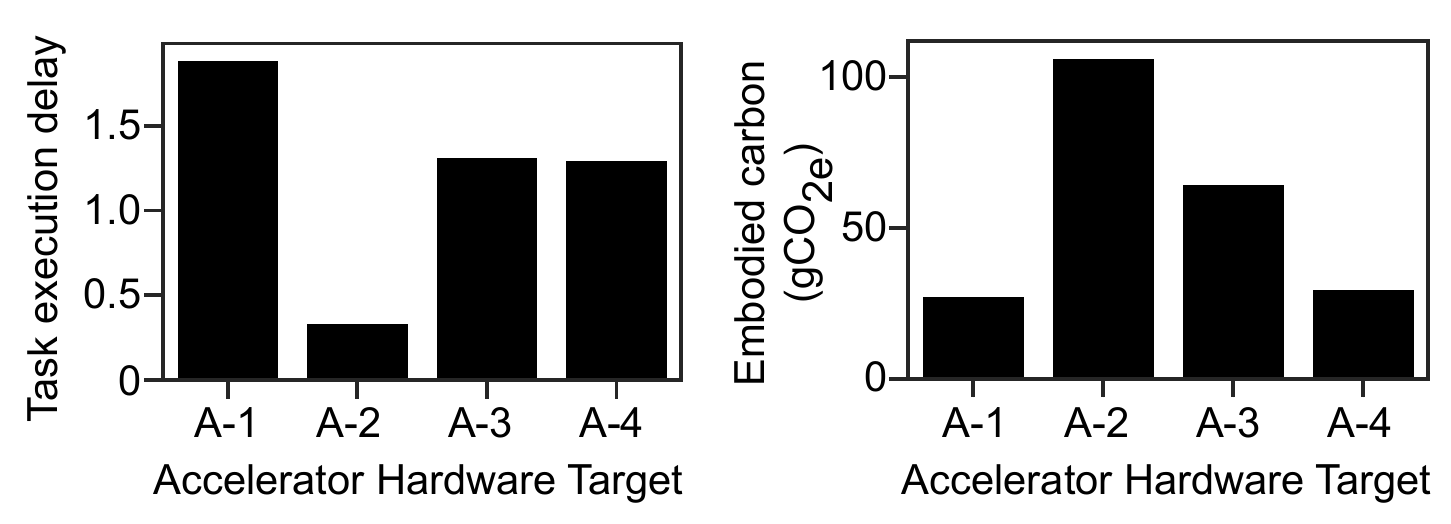}
\vspace{-0.4cm}
\caption{Total task delay (i.e. performance trend) for all AI and XR kernels presented in Table~\ref{tab:workloads} and the embodied carbon of each accelerator.}
\label{fig:Accelerator-bar}
\vspace{-1.1em}
\end{figure}

Based on our framework's hardware design space exploration and optimization results, we compare the carbon efficiency results of the framework's optimal accelerator designs using \cdp~to the optimal accelerator designs using non-carbon aware optimization such as EDP. Figure~\ref{fig:comparison_plot} presents the carbon efficiency results normalized to \underline{All} the kernels. Optimizing hardware design using our carbon-efficiency metric \cdp~compared to the design optimized for carbon-oblivious metrics, such as EDP, yields between 1.2 to 6.9 times better carbon efficiency improvement. 

\subsection{Designing for Specific Hardware Operational Lifetimes Yields Different Accelerators with Varying Carbon Efficiency}
\label{sec:results-ar-vr-accelerators}
\label{sec:results-3}

Based on our hardware design space exploration and optimization results for performance, power, area (PPA) and carbon, we produce four hardware accelerators---\textit{A-1} to \textit{A-4}. 
Figure~\ref{fig:Accelerator-bar} illustrates the latency performance and embodied carbon for the four hardware accelerators. 
Accelerator A-2 performs approximately 4 times better than A-3 and A-4, and 5.5 times better than A-1 (Figure~\ref{fig:Accelerator-bar}(a)). On the other hand, A-2 has the highest embodied carbon (Figure~\ref{fig:Accelerator-bar}(b)).

\begin{figure}[t!]
\centering
\includegraphics[width=0.96\columnwidth]{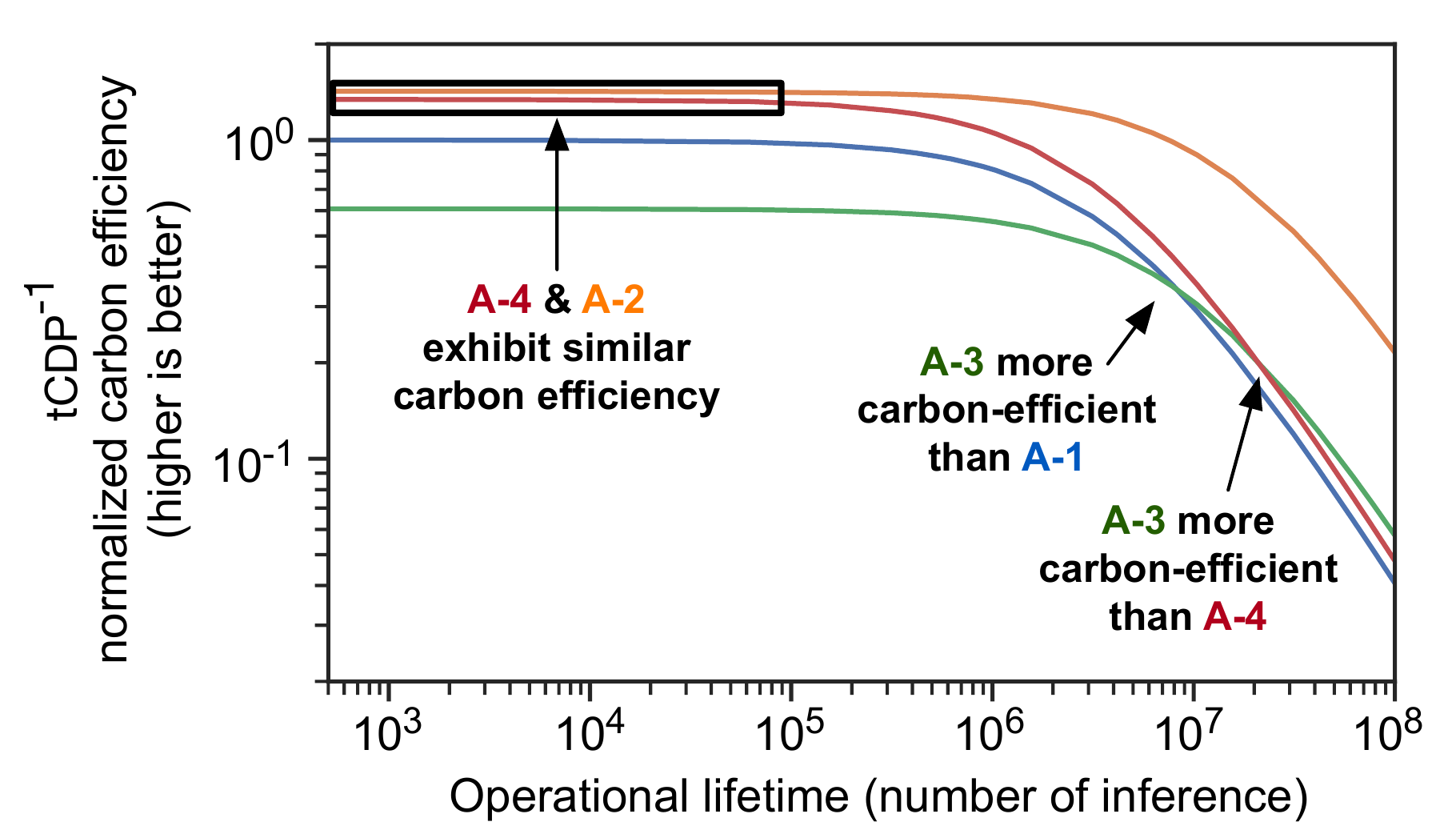}
\vspace{-0.3cm}
\caption{We demonstrate the carbon efficiency trade-offs (y-axis) between different accelerators running the same AI and XR workloads as the operational lifetime in number of inference (x-axis) varies.}
\label{fig:TCDPAccelerator}
\vspace{-1.1em}
\end{figure}

Figure~\ref{fig:TCDPAccelerator} illustrates the carbon efficiency results for the four accelerators (y-axis) for different operational lifetimes specified in number of inference (x-axis). The carbon efficiency curves are normalized to A-1 designed for $10^3$ inference: A-1 (blue), A-2 (orange), A-3 (green), and A-4 (red).

Carbon efficiency optimal hardware varies depending on the use duration the hardware is designed for. Here, the operational lifetime in units of number of inference determines the ratio of embodied and operational carbon of the system hardware.  
Figure~\ref{fig:TCDPAccelerator} shows, when shorter use phases are targeted at the hardware design time, i.e. number of inference below 10$^5$, accelerators A-2 and A-4 exhibit similar carbon efficiency results, despite accelerator A-4 having a significantly lower performance than A-2. This is because embodied carbon dominates operational carbon and accelerator A-4 has about 4 times lower embodied carbon. As operational exceeds embodied carbon, investing more embodied carbon in the hardware justifies the operational carbon efficiency and performance gains for accelerator A-2. 
Thus, as the operational lifetime of the hardware is prolonged from 10$^5$ to 10$^8$ inference, A-2 becomes significantly more carbon-efficient than A-4 due to A-2's significant performance and operational efficiency benefits.

Similar carbon efficiency inflection point is observed for accelerators A-1 and A-3, as well as, A-3 and A-4, respectively. In the range between 10$^6$--10$^7$ inference, A-3 shifts from 20\% operational carbon to 70\% operational carbon dominance, while A-1 shifts from 36\% to 85\% operational carbon dominance. Accordingly, the carbon-efficient design point switches from the low embodied carbon A-1 accelerator to the higher embodied carbon but higher performing and lower operational carbon A-3 accelerator. 
We also observe a carbon efficiency cross-over between accelerators A-4 and A-3, which exhibit similar task performance (within 1\% difference). As the usage of A-3 and A-4 become more operational carbon dominant, A-3 demonstrates higher carbon efficiency due to its lower operational energy outweighing A-4's lower embodied carbon.

\subsection{Optimizing Carbon Efficiency for General-Purpose System Hardware}
\label{sec:results-existing-arvr-oculus}

\begin{figure}[t!]
\includegraphics[width=\columnwidth]{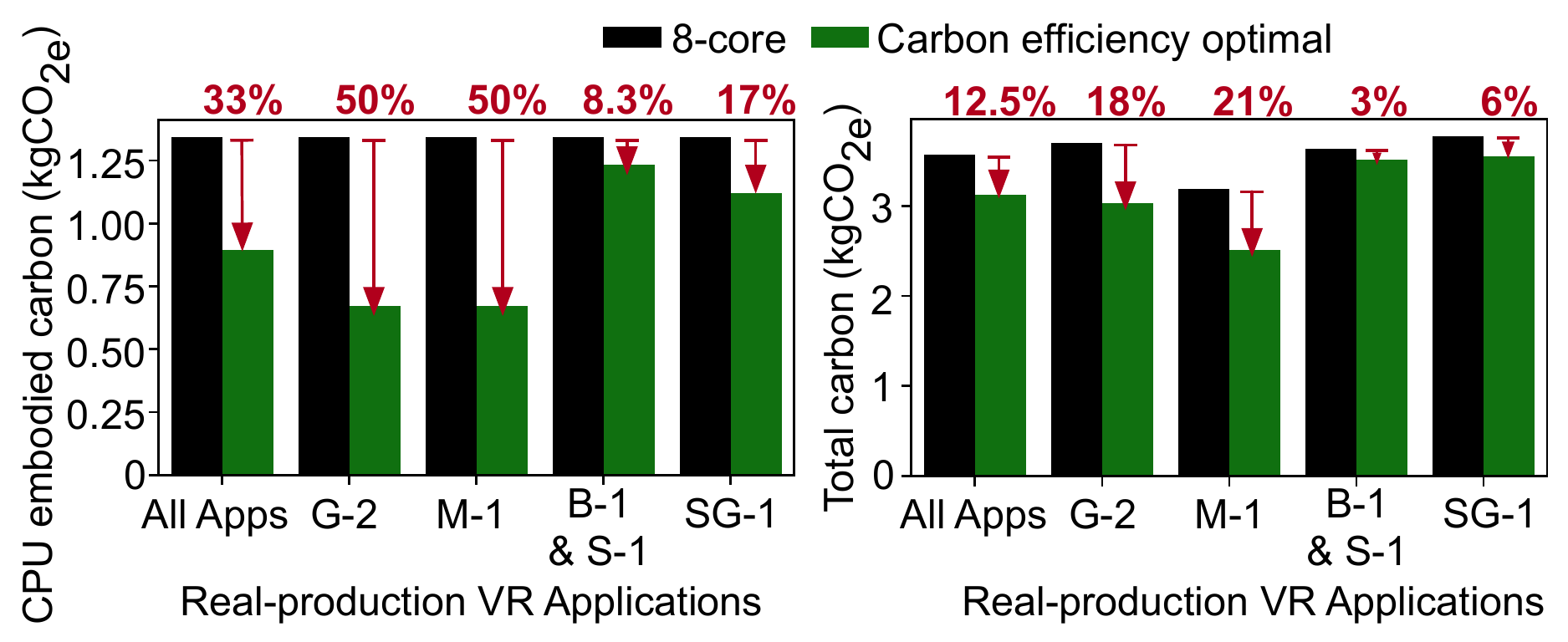}
\vspace{-0.65cm}
\caption{Embodied carbon and total carbon savings due to using hardware provisioning as a design knob when optimizing for the carbon efficiency of real-production VR CPUs for a set of applications.}
\label{fig:oculus-carbon}
\vspace{-1.1em}
\end{figure}

In addition to designing low-carbon hardware, we demonstrate carbon efficiency optimization opportunities for real-production VR devices running realistic XR applications. By provisioning hardware for application-specific characteristics, embodied carbon can be amortized more effectively. Below we quantify the potential of hardware utilization improvement.

Figure~\ref{fig:oculus-carbon} presents the embodied and life cycle carbon reduction results, leading to more carbon-efficient VR devices. Using our framework, we demonstrate up to 50\% embodied carbon savings when the CPU core count configuration is tailor-optimized for G-2 or M-2 applications. For top VR applications, an average of 33\% embodied carbon reduction can be realized by provisioning CPU core counts accordingly. 
Overall, the total life cycle carbon of presented VR systems can be improved by an average of 12.5\%. This is expected because turning off CPU cores leads to less significant operational carbon reduction; whereas reducing number of CPU cores results in more meaningful overall carbon reduction.

\begin{figure}[t!]
\centering
\includegraphics[width=0.85\columnwidth]{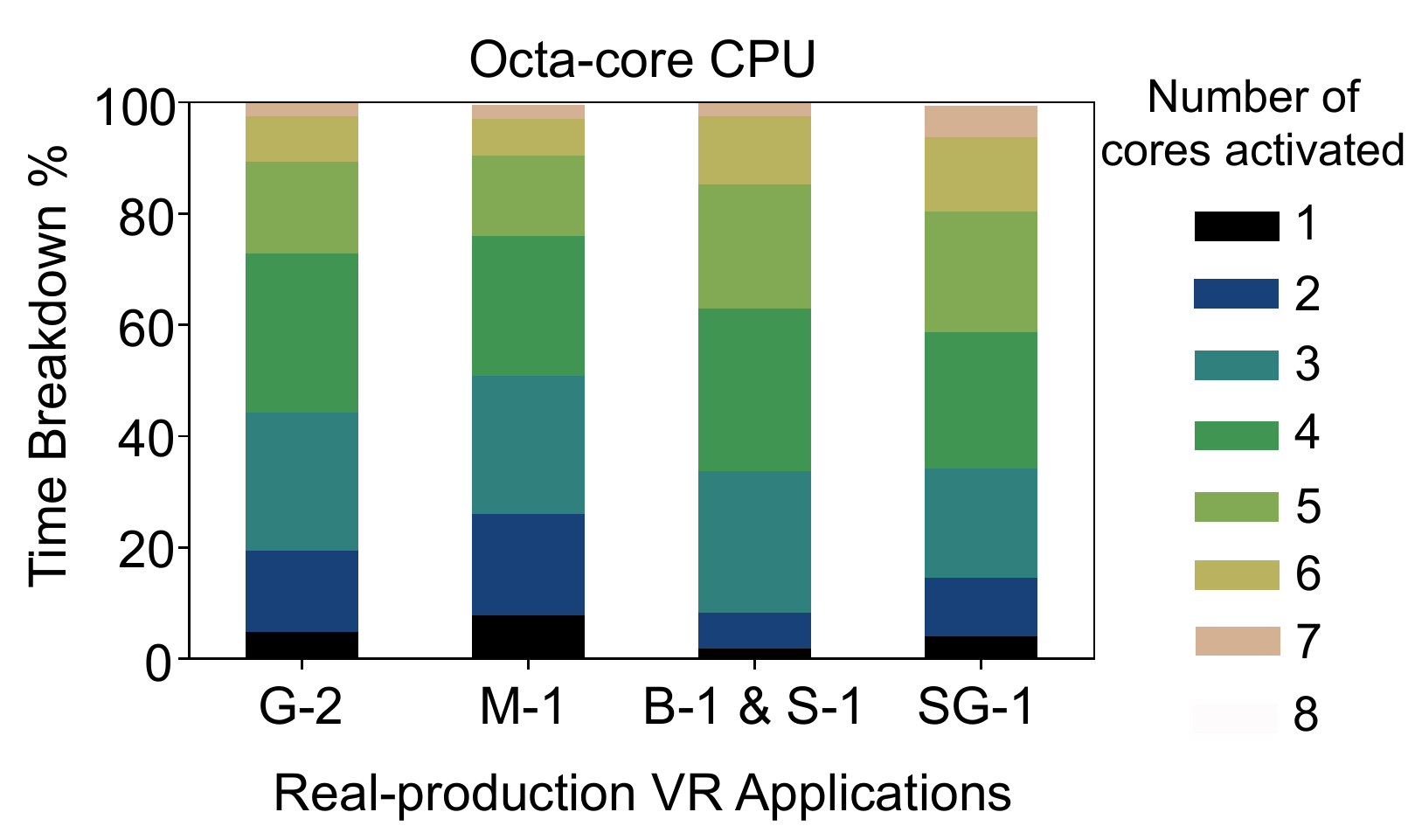}
\vspace{-0.4cm}
\caption{Time breakdown of number of cores activated by Android OS when using each application on a real-production Meta Quest 2 VR headset as per our profiling.}
\label{fig:tlp-time-breakdown-oculusapps}
\vspace{-1.1em}
\end{figure}

Taking a step further to illustrate the source of carbon efficiency improvement, Figure~\ref{fig:tlp-time-breakdown-oculusapps} quantifies the degree of thread-level parallelism (TLP)\footnote{TLP is the number of cores activated over the application execution time, such as: $\textrm{TLP} = \frac{\sum^{n}_{i=1} c_{i} i}{1 - c_{0}}$ ~\citep{Blake2010utilization,Flautner2000TLP}. $c_{i}$ is the fraction of time in which i CPU cores are concurrently running and n is the total number of cores.} for key real-production applications.
The x-axis represents the four VR applications (G-2, M-1, B-1 and S-1, SG-1) deployed on the Octa-core CPU.

Whereas the y-axis presents the time breakdown for the various TLP levels. Over the four representative VR applications, TLP ranges from 3.52 to 4.15 for the Octa-core CPU.

Compared with desktop and smartphone use cases of $\sim$2~\citep{Blake2010utilization} and 1.46~\citep{pandiyan:iiswc2013,mobileUtilization2015} average TLPs, modern VR applications exhibit higher hardware utilization parallelism with 3.9 average TLP. The core application kernels utilize three of the four gold cores on the Octa-core CPU. While auxiliary services, such as motion tracking, Inside-Out Tracking (IOT), and audio activities, utilize the four power-efficient silver cores.
There are at least three \textit{unused} cores at any point in time. The over-provisioned hardware presents embodied carbon reduction opportunities.

\begin{figure}[t!]
\centering
\includegraphics[width=\columnwidth]{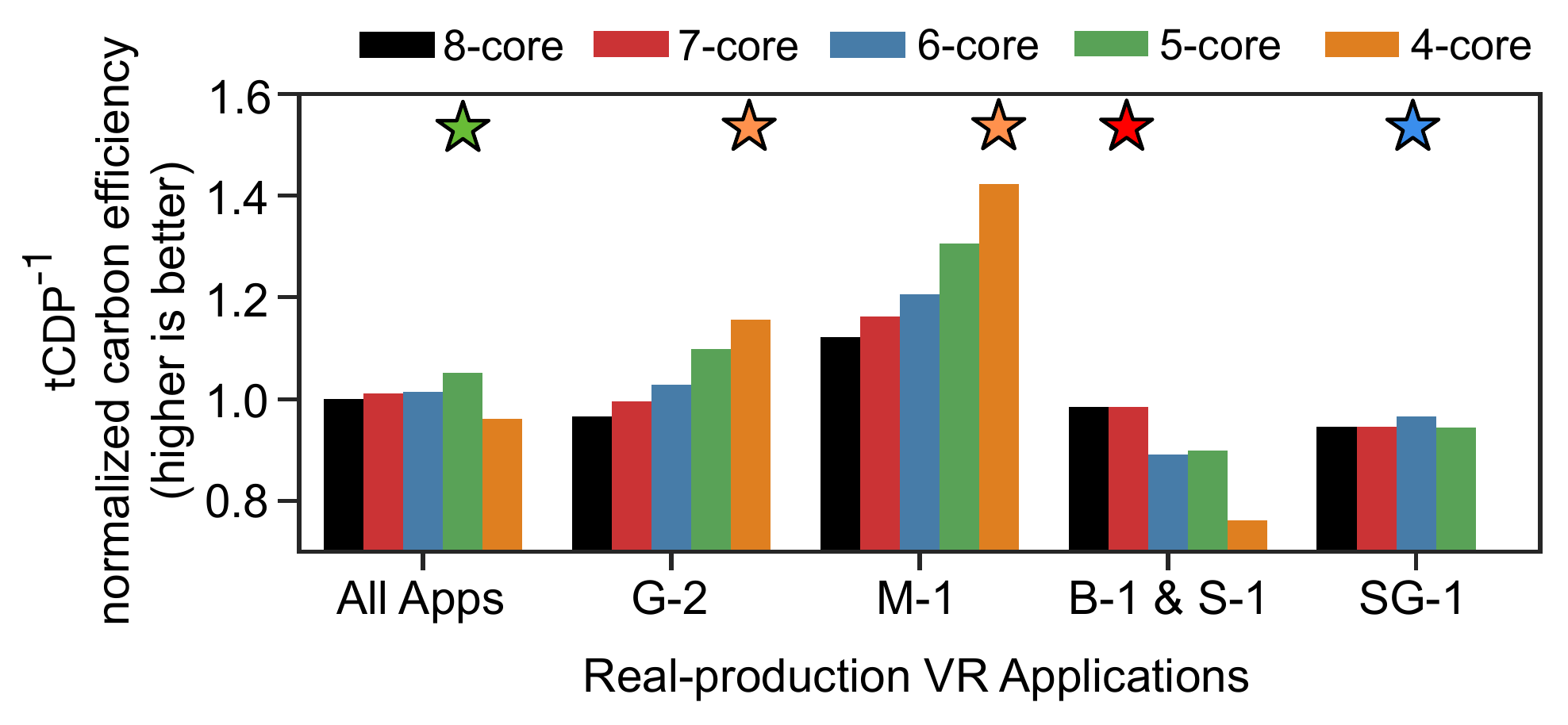}
\vspace{-0.8cm}
\caption{Carbon efficiency of different applications on a  production VR headset with varying CPU core-count configurations. Stars indicate the optimal carbon-efficient core configuration without sacrificing QoS or performance.}
\label{fig:oculus-carbon-efficiency}
\vspace{-1.1em}
\end{figure}

Across the top real-world VR applications, the TLP results indicate promising CPU core count reduction opportunities, leading to reduced embodied carbon with properly provisioned hardware and negligible performance penalty. While the room for per-device embodied carbon reduction may seem small, the impact over billions of VR devices deployed at scale can be significant.
Our framework enables computer architects and designers to design and optimize for more carbon-efficient hardware provisioned systems. 

In Figure~\ref{fig:oculus-carbon-efficiency}, using the matrix formalization, we present the carbon-efficient optimal core configuration for each of the VR application workloads (x-axis). The total task execution delay was computed as the reciprocal of the measured frame rate for each application using the different core configurations. The All Apps application workload indicates optimizing for the collective carbon efficiency of all applications--G-1, M-1, B-1 \& S-1, SG-1. We identify optimal carbon-efficient 5-core CPU configuration for All Apps, 4-core for G-2 and M-1, 7-core for B-1 \& S-1, and 6-core for SG-1. 

\subsection{Carbon-Efficient Hardware Replacement Frequency Varies with Operational Lifetime and Use}
\label{sec:rebuttal-hardware-lifetime-replacement}

\begin{figure}[t!]
\centering
\includegraphics[width=\columnwidth]{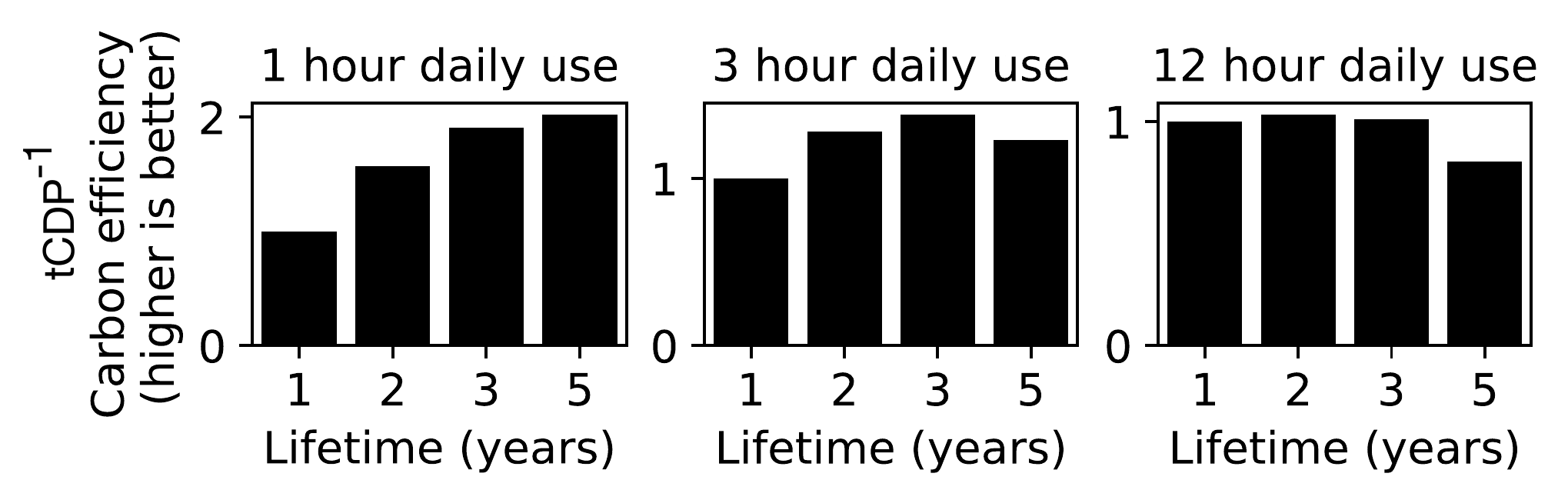}
\vspace{-0.8cm}
\caption{The optimal carbon-efficient hardware lifetime (normalized to 1-year) varies with operational lifetime and hardware energy efficiency.}
\label{fig:lifetime_hardware_replacement}
\vspace{-1.1em}
\end{figure}

We investigate the impact of hardware replacement frequency on the carbon efficiency of a real-production VR headset. Figure~\ref{fig:lifetime_hardware_replacement} illustrate how the most carbon-efficient lifetime for a hardware system varies with operational lifetime and usage. On the horizontal axis we vary the hardware lifetime; where one year represents frequent hardware replacements and five years represents infrequent refreshes. From left to right, we vary the operational use from one-hour to three-hour to twelve-hour daily. Using our framework, we compute the operational carbon based on 1.21$\times$ average annual energy efficiency improvement due to hardware advancements~\citep{ACTGupta2022}, and the TDP of Qualcomm Snapdragon SoC (refer to Figure~\ref{fig:oculus-apps-power-carbon}) of a real-production VR headset.

We observe that extending hardware lifetime significantly depends on operational use and the resulting interplay between operational and embodied carbon. For instance, for one-hour daily use, the most carbon-efficient hardware lifetime to design for is 5 years. This is because embodied carbon dominates operational carbon of the system. However, as the user's usage increases to three hours and twelve hours daily, the optimal carbon-efficient hardware lifetime shifts to 3 years and 2 years respectively. This is due to the increased device usage resulting in operational carbon dominance. Where frequent hardware replacements reap annual energy efficiency improvements, driving down the system's total life cycle carbon. For the example system, the framework's carbon savings are 50.5\% between optimal 5-year and 1-year lifetimes for 1-hour daily use, 27.5\% between optimal 3-year and 1-year lifetimes for 3-hour daily use, and 20.7\% between optimal 2-year and 5-year lifetimes for 12-hour daily use. This emphasizes the importance of extending software-hardware compatibility based on the minimum expected hardware usage and operational lifetime to reap system carbon efficiency benefits.

\subsection{Achieving Carbon Efficiency with Advanced 3D Integration Technologies}
\label{sec:results-5}

\begin{figure}[t]
\includegraphics[width=\columnwidth]{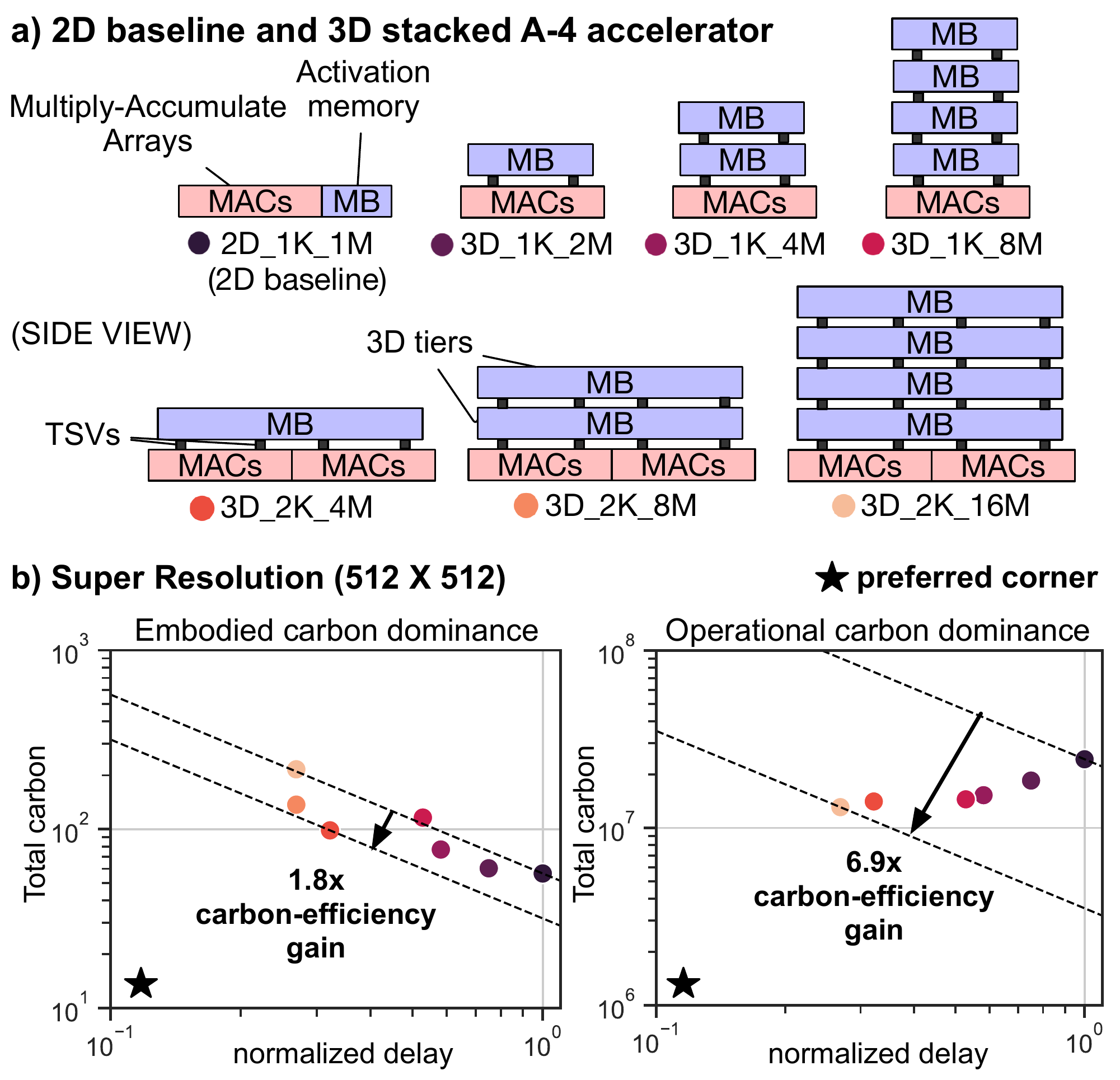}
\vspace{-0.75cm}
\caption{We explore the carbon efficiency gains for advanced 3D integration. (a) shows 2D baseline configuration of A-4 accelerator compared to six 3D memory stacking configurations in the z-direction~\citep{3darvrMeta}. (b) For SR(512$\times$512), 3D integration achieves 1.08$\times$ and 6.9$\times$ carbon efficiency gains, depending on the embodied to total life cycle carbon ratio, compared to the 2D baseline.}
\label{fig:3d-SR-kernels}
\vspace{-1.1em}
\end{figure}

Finally, we explore the role of emerging circuit integration technologies on carbon efficiency for future XR systems with stringent form factor requirements. Because of the tight die area limitation and yield implication, it is challenging to scale 2D integrated circuits (ICs) further. Additionally, 2D off-chip memory interfaces are prohibitively energy intensive and bandwidth limited for XR devices~\citep{3darvrMeta}.
 
We evaluate the carbon efficiency of 3D stacking, using face-to-face (F2F) bonding with hybrid bumps, for hardware accelerator A-4 tailor-designed for select XR kernels in Table~\ref{tab:workloads}. Figure~\ref{fig:3d-SR-kernels}(a) shows the 2D baseline SoC memory disaggregation architecture compared to different 3D memory-stacked configurations, offering high-density vertical integration to local memory and compute. K and M represent number of MAC arrays and on-chip SRAM capacity respectively.

The embodied carbon computation for 3D stacking only takes into account the stacked dies. The carbon cost of the high-density 3D through-silicon-vias (TSVs) and the manufacturing process of stacking are not included due to lack of data. Figure~\ref{fig:3d-SR-kernels}(b) presents the carbon efficiency improvement of 3D accelerators designed for SR (512$\times$512) for 80\% (left) and 6\% (right) embodied carbon to total life cycle carbon. In embodied carbon dominance case, our framework achieves 1.8 times carbon efficiency improvement using 3D stacked 2K MACs and 4MB SRAM (3D\_2K\_4M). In operational carbon dominance case, our framework achieves 6.9 times carbon efficiency improvement using 3D stacked 2K MACs and 16MB SRAM (3D\_2K\_16M).

\begin{figure}[t]
\centering
\includegraphics[width=1.03\columnwidth]{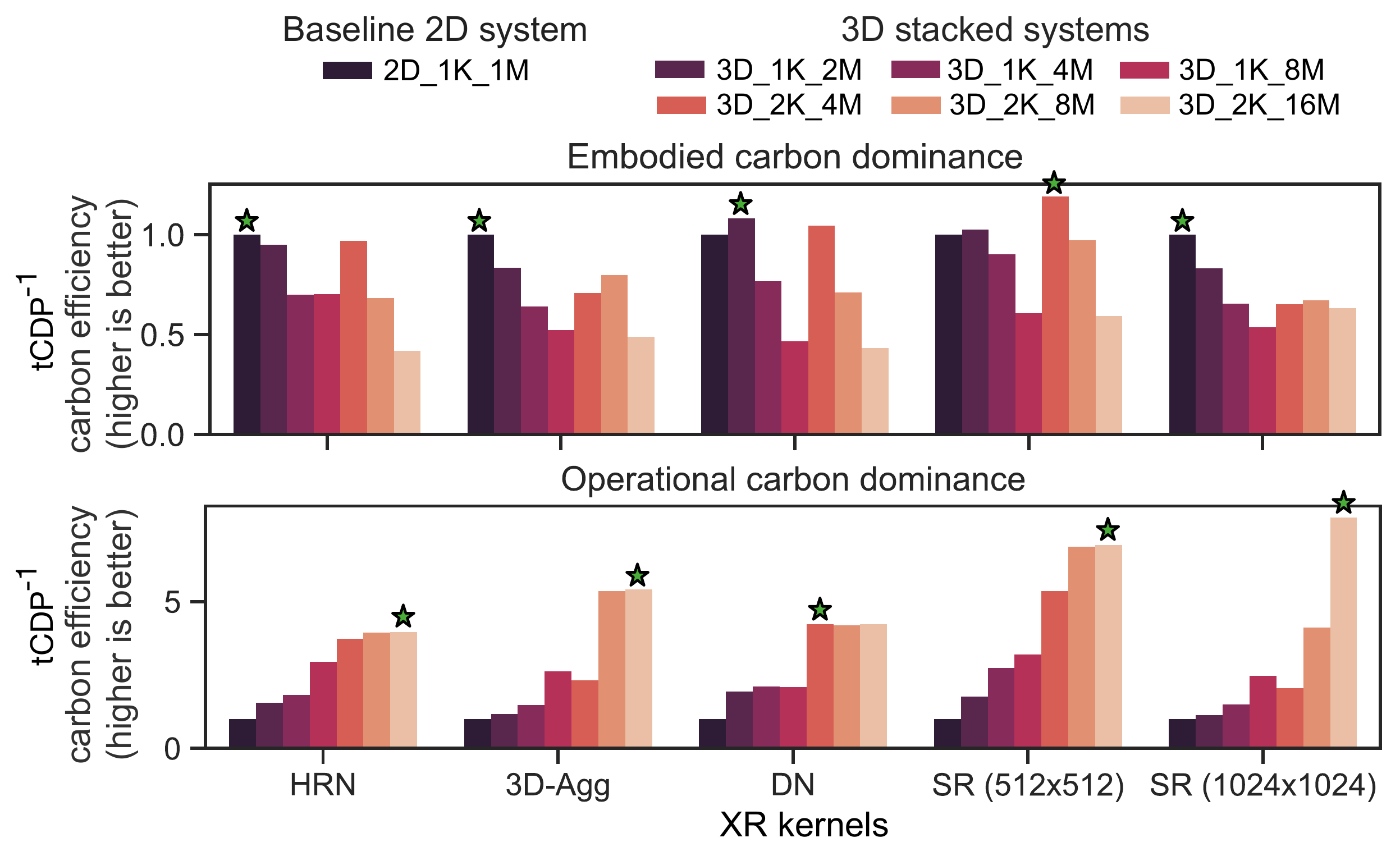}
\vspace{-0.75cm}
\caption{Carbon efficiency of 3D stacked IC configurations normalized to 2D baseline per XR kernel (stars indicate optimal). 3D stacking shows limited carbon efficiency benefits in the 98\% embodied to total carbon case (top). In the 6\% embodied to total carbon case (bottom), 3D stacking can reap up to 7.9$\times$ carbon efficiency over the baseline.}
\label{fig:3d-bar}
\vspace{-1.1em}
\end{figure}

Figure~\ref{fig:3d-bar} shows the carbon efficiency results of 3D stacked configurations normalized to the 2D baseline. In embodied carbon dominance case, 2D baseline is the most carbon-efficient configuration for HRN, 3D-Agg and SR (1024$\times$1024). We observe limited carbon efficiency benefits for some 3D configurations; for SR (512$\times$512), 3D\_2K\_4M due to higher compute, and for DN, 3D\_1K\_4M due to higher SRAM. 
When operational carbon dominates use, the carbon-efficient optimal shifts towards 3D stacked ICs. 3D\_2K\_8M and 3D\_2K\_16M demonstrate similar carbon efficiency benefits of 4, 4.2, 6.9 times for HRN, 3D-Agg, and SR (512$\times$512), respectively. 
For compute and memory intensive SR (1024$\times$1024), 3D\_2K\_16M improves carbon efficiency 7.86 times.  
\section{Related Work}
Prior efforts have characterized and investigated the rising environmental impact of different computing systems~\citep{chasingGupta2021, Freitag2021, imec2020, imec2022logic, totallygreen2012hplabs, challenges2002JainWullert}.
To tackle this growing problem, a variety of carbon modeling tools and methodologies have been proposed to quantify and evaluate computing systems' carbon footprint~\citep{ Greenchip2019, ACTGupta2022, iot2022lifecyclemodeling, boyd2011thesis}.
These models enable embodied and operational carbon accounting and consider sustainability as a first-order design metric. Across the proposed solutions, sustainability-aware carbon optimization yield distinct designs compared to optimizing for performance, power, and area. 
More recent work proposed a first-order model considering both embodied and operational carbon proxies to overcome current uncertainties in carbon data~\citep{eeckhout2022}. While these models provide early-stage insights into the environmental footprint of systems, they lack the framework and metrics to enable carbon-efficient system optimization and design. In this work, we demonstrate how to design and optimize carbon-efficient computing systems, \textit{accounting for performance, power, area, and total life cycle carbon simultaneously}.
\section{Conclusion}
The primary goal of this work is to enable a path towards designing low carbon computing systems without sacrificing performance nor operational efficiency. First, we demonstrate that carbon-aware metrics should capture total life cycle carbon and characterize the degree of \textit{unused} embodied carbon due to over-provisioning in existing hardware systems. Next, we propose a closed-loop cross-stack hardware optimization framework that yields carbon-efficient computing systems. We introduce carbon efficiency metric, \cdp,~that captures the total life cycle carbon, including the changing relative ratio between embodied \textit{and} operational carbon. Finally, we use our framework to design and optimize hardware accelerators and general purpose system hardware for a variety of XR workloads and applications. We hope this work lays the foundation for carbon-efficient design and optimization across the computing stack and paves the way for sustainable computing forward.    

\section*{Acknowledgements}
We would like to thank the colleagues at Meta: Jordan Tse, Noah VanGorder, Lita Yang, Edith Beigne for many valuable discussions and feedback on the work.     

\bibliographystyle{ACM-Reference-Format}
\bibliography{refs}

\end{document}